\documentclass[manuscript,screen]{acmart}
\usepackage{makecell,rotating,multirow,diagbox}
\usepackage{caption}
\usepackage{subfigure}
\usepackage{indentfirst}
\AtBeginDocument{%
  \providecommand\BibTeX{{%
    \normalfont B\kern-0.5em{\scshape i\kern-0.25em b}\kern-0.8em\TeX}}}
\begin{document}
\title{Utilizing BERT for Information Retrieval: Survey, Applications, Resources, and Challenges}
 \author{Jiajia Wang}
\affiliation{%
  \institution{School of Sciences, Henan University of Technology}
  \city{Zhengzhou, Henan 450001}
  \country{China}
}
\author{Jimmy X. Huang}
\affiliation{%
  \institution{   Information Retrieval and Knowledge Management Research Lab, York University}
  \city{Toronto}
  \country{Canada}
}
\authornote{Corresponding author: Jimmy X. Huang (jhuang@yorku.ca)}

\author{Xinhui Tu}
\affiliation{%
  \institution{School of Computer Science, Central China Normal University}
  \city{Wuhan}
  \country{China}
}
\author{Junmei Wang}
\affiliation{%
  \institution{School of Computer, Hangzhou Dianzi University}
  \city{Hangzhou}
  \country{China}
}
\author{Angela J. Huang}
\affiliation{%
  \institution{Lassonde School of Engineering, York University }
  \city{Toronto}
  \country{Canada}
}
\author{Md Tahmid Rahman Laskar}
\affiliation{%
  \institution{York University \& Dialpad Inc.}
  \city{Toronto}
  \country{Canada}
}
\author{Amran Bhuiyan}
\affiliation{%
  \institution{Information Retrieval and Knowledge Management Research Lab, York University}
  \city{Toronto}
  \country{Canada}
}
\renewcommand{\shortauthors}{J. Wang, J. X. Huang, X. Tu, J. Wang, A. J. Huang, M. T. R. Laskar and A. Bhuiyan}
\begin{abstract}
Recent years have witnessed a substantial increase in the use of deep learning to solve various natural language processing (NLP) problems.
Early deep learning models were constrained by their sequential or unidirectional nature, such that they struggled to capture the contextual relationships across text inputs.
The introduction of bidirectional encoder representations from transformers (BERT) leads to a robust encoder for the transformer model that can understand the broader context and deliver state-of-the-art performance across various NLP tasks. This has inspired researchers and practitioners to apply BERT to practical problems, such as information retrieval (IR). A survey that focuses on a comprehensive analysis of prevalent approaches that apply pretrained transformer encoders like BERT to IR can thus be useful for academia and the industry. In light of this, we revisit a variety of BERT-based methods in this survey, cover a wide range of techniques of IR, and group them into six high-level categories: (i) handling long documents, (ii) integrating semantic information, (iii) balancing effectiveness and efficiency, (iv) predicting the weights of terms, (v) query expansion, and (vi) document expansion. We also provide links to resources, including datasets and toolkits, for BERT-based IR systems. A key highlight of our survey is the comparison between BERT's encoder-based models and the latest generative Large Language Models (LLMs), such as ChatGPT, which rely on decoders.  Despite the popularity of LLMs, we find that for specific tasks, finely tuned BERT encoders still outperform, and at a lower deployment cost.  Finally, we summarize the comprehensive outcomes of the survey and suggest directions for future research in the area. 
\end{abstract}
\vspace{-50cm}
\ccsdesc[500]{Information systems}
\ccsdesc[500]{Information systems~Information retrieval}
\keywords{BERT, Information Retrieval, Natural Language Processing, Artificial Intelligence}
\maketitle
\section{Introduction and Motivation}
Information retrieval (IR) is the acquisition of informational resources from a large volume of unstructured
information that satisfies the user's needs \cite{schutze2008introduction}. Web search engines are a typical example of an information retrieval system. IR models are used to sort and rank the retrieved documents according to the user's queries, this is essential for designing search engines. Traditional IR models include the $vector\ space\ model$, the $probability\ model$, and the $language\ model$. In 1959, Arthur Samuel introduced the term $machine\ learning$ as a method of data analysis that automates the building of an analytical model. Machine learning is a branch of Artificial Intelligence (AI) based on the idea that the system can learn from data, identify patterns, and make decisions with minimal human intervention.

Learning to rank (LTR) is an approach that involves utilizing machine learning techniques to construct ranking models. Unlike traditional and manually constructed ranking formulae, LTR uses data to automatically build a model that can make predictions without being explicitly programmed to do so. While it delivers superior performance to traditional IR models, its capabilities can be further enhanced in two areas. First, the complexity and diversity of languages and cultures make it challenging to accurately comprehend semantic information embedded within queries and document content. Second, feature engineering is manually performed in machine learning, and requires input from domain experts, where this incurs a significant cost.

Deep learning is a domain of machine learning that uses artificial neural networks with multiple layers to learn patterns from the given data. It has significantly outperformed traditional machine learning models in recent years to become the focus of AI research. Deep learning has exhibited an impressive capability in a wide range of tasks, such as speech recognition \cite{Hin}, natural language processing \cite{surveynlp}, recommender systems \cite{surveyrecommender},  and computer vision \cite{Kri}. For more information on deep neural networks, we refer the interested reader to \cite{Wei} and \cite{dcn}.

In early research in the area, vectors of word embedding like word2vec\footnote{\url{https://radimrehurek.com/gensim/models/word2vec.html}} \cite{word2vec}
and GloVe\footnote{\url{https://nlp.stanford.edu/projects/glove/}}\cite{glove} were used in neural network models to obtain the representation
of the words in the input texts. Word2vec uses one of the following two models to this end:
the (i) continuous bag-of-words (CBOW) model or the (ii) continuous skip-gram model. The CBOW model predicts a given word according to the words surrounding it within a fixed window, where the order of the surrounding words does not impact the results of its predictions.
By contrast, a continuous skip-gram predicts the surrounding words in a fixed window based on the given word. This causes the nearby contextual words to have a more significant influence on the predicted word than the contextual words that are more distant. GloVe is a tool for word representation based on global term frequency. The word representations generated by GloVe capture
some semantic information between words. However, neither word2vec nor GloVe can adequately capture contextual
information. Contextualized information refers to the intelligence within the given discourse that exploits the relationships between the words, phrases, sentences, and even paragraphs by considering all the circumstances (e.g., bidirectional contexts) involved in the emergence of the text. For example, the word \textit{bank}  appears in \textit{bank account} and \textit{bank of the river}. If a model generates the same representation for \textit{bank} in both scenarios, this means that it does not capture the contextual information. A context-based model can generate different representations for \textit{bank} based on the words preceding and following it (i.e., bidirectional contextual information) in the sentence.

Considering the importance of bidirectional contextual information, Devlin et al. proposed Bidirectional Encoder Representations from Transformers (BERT) \cite{Devl} in 2018. It leverages the transformer encoder \cite{Ash} to pretrain a bidirectional language model. BERT has demonstrated an impressive capability in terms of understanding language in various NLP tasks: (i) It outperformed previously proposed language models \cite{peter,GPT} on the GLUE \cite{GLUE} benchmark, and delivered state-of-the-art performance on tasks like natural language inference, textual entailment, sentiment analysis, and sentence similarity modeling. (ii) It demonstrated impressive comprehension on the SQuAD v$1.1$ test set by outperforming even humans on two measures of evaluation (ExactMatch (EM) and the F$1$ score). (iii) BERT has a good capacity for transfer learning \cite{Dom, Lin,laskar20} that helps it adapt to different tasks through pretraining or fine-tuning.

The successful use of BERT in various applications involving language comprehension has also inspired researchers to apply BERT-based approaches to IR ~\cite{laskarlrec,pan2022,Mallia2021,Formal2021,Hofsttter2021}. In light of the recent successes of BERT-based models in IR, and given the significant number of studies in this domain, we provide a survey in this study that brings together a number of BERT-based approaches into a coherent resource that provides a single point reference for research on IR. Note that few surveys~\cite{Guo,mitra,onal} in the literature have highlighted neural approaches to IR. For instance, Onal et al. \cite{onal} focused on the fundamentals of text information retrieval, while Mitra et al. \cite{mitra} presented a brief review of IR methods that use pretrained embeddings and neural networks. Closer to our work here, Guo et al. \cite{Guo} surveyed IR approaches that focus only on ranking models using deep neural networks. Unlike these approaches, our survey targets IR approaches that utilize pretrained transformer encoders like BERT and its variants while also systematically reviewing the challenges of using BERT-based models for IR.

Since 2022, there has been remarkable progress in leveraging Language Models (LMs), encompassing Pretrained Language Models (PLMs) and Large Language Models (LLMs). PLMs such as BERT~\cite{Devl}, RoBERTa~\cite{RoBERTa}, and GPT \cite{GPT} undergo pretraining on extensive textual corpora to assimilate worldly knowledge. The recent advent of LLMs, exemplified by OpenAI’s ChatGPT \cite{laskaracl} and GPT-4 \cite{gpt4}, has catalyzed an unforeseen surge in innovation, highlighting the diverse possibilities of AI.

While decoder-only LLMs like ChatGPT\footnote{\url{https://chat.openai.com/}} \cite{laskaracl} have gained significant attention for
their ability to generate human-like text, there remain many challenges in using LLMs in real-world problems. For instance, they require significant computational resources for both training and inference. In addition, they can generate responses based only on their pretraining knowledge, but cannot answer any time-sensitive questions that are beyond their pretraining timeline. Moreover, using LLMs as a service by utilizing their APIs poses privacy-related concerns. Considering their cost, failure to adapt over time, and privacy-related issues, it is challenging to use such LLMs in real-world applications. As an alternative, encoder-based transformer models like BERT do not pose privacy-related risks and require significantly lower computational resources. BERT-based models are particularly adept at information retrieval tasks \cite{jimmylinacl} that are crucial for applications like search engines and recommendation systems owing to their effective semantic understanding, low computational requirements, and efficient response times. BERT-based rich contextual embeddings are useful for retrieving information from external resources to augment the context of the LLM during inference \cite{Qia}. In this regard, we aim to review the use of pretrained transformer encoder models like BERT in IR. This can help fill a gap in the relevant research, and can provide a comprehensive overview for researchers and industry professionals.

The major contributions of this survey in this context are four-fold. First, we logically organize the aspects of BERT for ad-hoc IR in terms of the updated literature. This enables the reader to understand BERT-related models from various perspectives in a step-by-step manner and discover new research opportunities by following this organization. Second, we analyze the advantages and disadvantages of different models based on BERT, and conduct a number of state-of-the-art comparisons in different categories. Third, we present the challenges to the use of BERT for IR and examine important IR toolkits. Finally, we provide comprehensive conclusions based on the outcomes of the survey and outline the trend of future research on BERT-based IR.

The remainder of this paper is organized as follows: We discuss the background and structure of BERT in Section~\ref{sec:back}. Section~\ref{sec:application} discusses the use of BERT for ad-hoc IR, including utilizing BERT for handling long documents, utilizing BERT for integrating semantic information, BERT for balancing effectiveness and efficiency, BERT-based term weight prediction, BERT-based query expansion, and BERT-based document expansion, respectively. In Section~\ref{sec:Res}, we present resources that incorporate datasets as well as BERT-based ranking models. Finally, we provide the conclusions of this survey in Section~\ref{sec:conclusion}.

\section{Background}
\label{sec:back}
The popularization of the Web and the explosive growth of Internet-based information have led researchers to focus on the performance and efficiency of IR systems as well as the technology of their core search engines. In addition to crawling and indexing in the pipeline, the most important factor while designing a search engine is ranking, which relies solely on IR models for sorting the relevant documents with respect to a user's query. Thus, the efficiency of a search engine heavily relies on the efficiency of the IR model.

Relevance-based ranking is the main problem in IR that plays a fundamental role in various applications of it, such as search engines. To satisfy the user's informational needs (queries), we need to identify the collection of documents that is the most relevant to their queries. Given a set of queries $Q=\{q_1,q_2,\ldots,q_m\}$ and a set of documents $D=\{d_1,d_2,\ldots,d_n\}$,
we need to use an IR model to rank the relevant documents based on search queries.

In this section, we discuss the most popular traditional IR models in Section~\ref{sec:traditional}, neural ranking models In Section~\ref{sec:neural}, pretrained language models in Section~\ref{sec:pretrained}, and improvements and extensions of pretrained language models in Section~\ref{sec:improvements}.

\subsection{Traditional Retrieval Models}
\label{sec:traditional}
Traditional retrieval models can be divided into four categories: Boolean models, vector space models, language models, and probability models. The first three models assume that documents and queries are composed of a set of words and ignore the positional information of tokens in a given query/document.

The Boolean model~\cite{salton, Lashkari} is a simple retrieval model based on set theory and Boolean algebra. It aims to find documents that return ``1" with a query term. However, it lacks the concept of document rank, which limits its retrieval capabilities. Salton et al. introduced a non-binary weight model called the $vector\ space\ model$ (VSM) \cite{Ton} to rank the relevant documents based on search queries. The VSM \cite{Ton} transforms text processing into vector computation in vector space, presents semantic similarity through spatial similarity, and thus compensates for the Boolean model's inability to scale the returned documents. However, approaches based on the VSM model fail to recognize semantic ambiguity in the text due to their dependence on token frequency, and cannot solve the problems of polysemy and synonymy. The BM25 probabilistic model \cite{Rob} is among the best weighting models for IR \cite{He2011,hiem, Huang2003, Huang2005, Huang2009,Zhao2011, Zhao2014,Yin2010,Chen2017,Liu2007}. It assigns a weight to a search term based on its intra-document term frequency and query term frequency. The document weight is the sum of the weights of the query term. Ponte et al. \cite{Lang} proposed a language model that considers a passage of natural language text as a discrete time series. They utilized probabilistic models to estimate the probability of a document being relevant based directly on a particular probability distribution and a language model to predict the probability of a given sequence of words occurring in a sentence.
\subsection{Neural Ranking Models}
\label{sec:neural}
Traditional retrieval models obtain ranking scores by searching for exact matches of words in both the query and the document. However, this limits the availability of positional and semantic information and may lead to $vocabulary$ $mismatch$ \cite{Cro}. By contrast, neural ranking models construct query-to-document relevance structures to learn feature representations automatically via word2vec \cite{word2vec} and GloVe \cite{glove}. There are two main classes of neural ranking models: representation-focused models and interaction-focused models.

$\bullet$ $\mathbf{Representation}\bf{-}\bf{focused\ Architecture}$:
This architecture encodes the query and the document separately. Thus, no interaction occurs between the query and the document during the encoding procedure. Following this pipeline, Huang et al. introduced the deep structured semantic model (DSSM) \cite{Dss} that can predict the semantic similarity between sentences and obtain a lower-dimensional semantic vector representation of a sentence. Similarly, Hu et al. in 2014 studied two models for text matching \cite{Cnn}, \textit{Architecture I (ARC-I)} and \textit{ARC-II}. The \textit{ARC-I} model uses a convolutional neural network (CNN) to separately obtain features from texts and calculate the similarities between the features. However, there is no interaction between the texts while the features are extracted.

$\bullet$ $\bf{Interaction\bf{-}focused\ Architecture}$:
This architecture evaluates the relationship between a query and a document according to their interaction. Compared with representation-focused architecture, interaction-focused architecture can capture more contextualized information. Hu et al. proposed the $\rm ARC-II$ model \cite{Cnn}, which uses a 1D convolutional layer to explore the interaction between input texts. The $\rm DRMM$ \cite{drmm}, introduced by Guo et al., uses a histogram to map the relevance between the query and words in the document. Wan et al. proposed $\rm Match\ SRNN$ \cite{Srnn}, which models complex interactions between input texts by using a neural tensor layer. Xiong et al. proposed $\rm KNRM$ \cite{Knr}, which uses a kernel pooling layer to calculate the relevance between the query and words in the document. Dai et al. introduced $\rm Conv\mbox{-}KNRM$ \cite{Zhu}, a model for the soft matching of n-grams for ad-hoc IR by using a CNN. Although all these approaches improve performance over a range of IR applications, they fail to consistently work well when trained on small-scale data.

\subsection{Pretrained Language Models}
\label{sec:pretrained}
Creating a better language model is crucial to NLP. In the broad sense, a pretrained language model refers to a language model trained on large-scale data, including distributed static representation-based models (word2vec \cite{word2vec} and GloVe~\cite{glove}) and dynamic contextualized models (CoVe \cite{CoVe} and ELMo \cite{peter}). Following the new wave of deep transformer-focused pretrained models \cite{GPT}, such as the generative pretrained transformer (GPT) \cite{GPT} developed by OpenAI, a new understanding of the pretrained language model, known as BERT \cite{Devl}, emerged in the NLP community in 2018. Compared with previous models~\cite{Cnn,Dss,Knr,Zhu}, pretrained models based on deep transformer encoders have the advantage of being able to process input text sequences in parallel. In addition, transformer encoder-based pretrained language models can focus on the target task and fine-tune it to achieve better performance than the previously mentioned models ~\cite{Cnn,Dss,Knr,Zhu}, where the representation of word vectors remains constant when used as an additional input to perform the target task.

\begin{figure}[t!]
 \centering
 \includegraphics[width=1.0\textwidth]{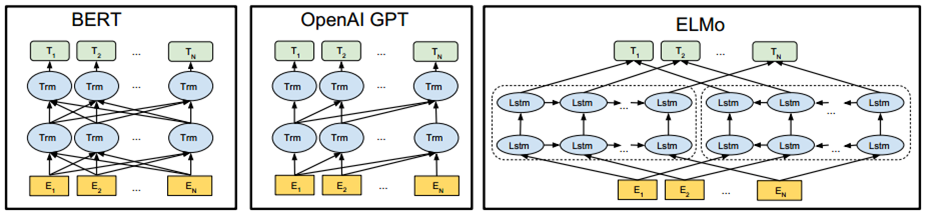}
\caption{The pretraining model architectures of BERT, OpenAI GPT, and ELMo. BERT demonstrates a significant bidirectional nature, OpenAI’s GPT exhibits a unidirectional characteristic, while ELMo uses a shallow bidirectional approach.}
 \label{f1}
\end{figure}

A comparison of pretrained transformer models reveals that the GPT model from OpenAI uses a multi-layer decoder based on a transformer. It is mainly composed of two parts: generative pretraining and differential fine-tuning. Unlike the GPT, BERT does not leverage the decoder of the transformer for self-regression language modeling. Instead, it introduces the auto-encoding pretraining task. In contrast to ELMo, which uses the left-to-right and right-to-left LSTM models, BERT uses the bidirectional transformer encoder as the language model, and takes the masked language model (MLM) and next sentence prediction (NSP) as the objectives of pretraining. Figure~\ref{f1} shows a comparison of ELMo, OpenAI GPT, and BERT.

\subsection{Improvements to and Extensions of Pretrained Language Models}
\label{sec:improvements}
The advantage of BERT’s auto-encoding language model is that it can use contextualized information on the given word to better predict its vector representation. Because BERT has pioneered a new area of NLP, many variants and extensions based on it have emerged.

Yang et al. introduced XLNET \cite{XLNET} in 2019. It is an auto-regressive language model based on Transformer-XL \cite{TranXL}, in which all tokens are predicted in random order, rather than sequentially, and the joint probability of each predicted token is factorized by using the product rule to address the issue of a discrepancy between pretraining and fine-tuning in BERT due to the use of artificial symbols like [MASK]. In addition, XLNET incorporates the advantages of the auto-encoding language model, and obtains contextualized information by introducing the permutation language model and two-stream self-attention. Shortly after XLNET overtook BERT across the board, Facebook proposed RoBERTa \cite{RoBERTa}, which stands for the Robustly Optimized BERT Pretraining Approach, which can achieve state-of-the-art performance on multiple
tasks. RoBERTa does not change the structure of Google's BERT but only its method of pretraining. The changes in the pretraining methods are mainly reflected in the following aspects: (1) Static masking vs. dynamic masking: In static masking, the masking operation is performed only once during data processing while in dynamic masking, the pattern of masking changes whenever a sequence is fed into the model. (2) With Next Sentence Prediction (NSP)  vs. without NSP (Removing the NSP from  RoBERTa's training procedure enables the input to be packed with full sentences sampled from different documents until the maximum length of the sequence is 512; this is called FULL-SENTENCES). (3) A larger mini-batch. (4) More training data and a longer training time.

As with BERT,  XLNET and RoBERTa encounter the issue of consistent growth in the size of the pretrained language model that leads to memory-related constraints, a longer preparation time, and unexpectedly poor performance. To reduce memory consumption and improve the speed of training of the model, Lan et al.~\cite{Albert} proposed ALBERT, which incorporates two parameter-reduction techniques: word vector decomposition and cross-layer parameter sharing. Pretraining in XLNET and RoBERTa by using NSP is simple and thus cannot be used to learn deep semantic information. ALBERT replaces it with sentence-order prediction (SOP) to predict if the sentences are coherent.
Different from the above-mentioned pretrained models, Clark et al. proposed ELECTRA \cite{ELECTRA} that combines BERT with a structure similar to that of the generative adversarial network (GAN) \cite{GAN}. ELECTRA consists of a generator and a discriminator and can outperform BERT while using fewer parameters and data to achieve results comparable to those of the SOTA model RoBERTa while using only a quarter of the number of calculations.

To test the validity of pretraining models across languages, Cui et al. presented MacBERT \cite{MacBERT} (MLM as a correction) model as a pretrained model. The model improves RoBERTa from several aspects, such as reducing the gap between the pretraining and the fine-tuning phases by masking similar words. It also replaces the NSP pretraining task with the SOP in ALBERT. In addition, BERT has been pretrained on biomedical \cite{biobert} and clinical \cite{clinicalbert} data to achieve state-of-the-art performance across biomedical tasks \cite{bionlp}.

To balance efficiency and effectiveness, Hofst\"{a}tter et al. proposed the transformer--kernel (TK) model \cite{Hofsttter2019, Hofsttter2020} that improves the architecture of the KNRM \cite{Knr}. The TK model encodes queries and documents independently by using a small number of transformer layers (up to three) before kernel pooling to obtain an optimal ratio between effectiveness and efficiency. Hofst\"{a}tter et al. also proposed the TK model with local attention (TKL) \cite{Hofsttter2020a}, which uses local self-attention rather than the self-attention of the encoder of the transformer to perform pooling over windows of document terms via a modified KNRM.

The above-mentioned pretrained language models require different training and fine-tuning phases to adapt to different tasks. By contrast, Raffel et al. introduced a unified architecture, T5 (text-to-text transfer transformer) \cite{Raffel2019}, that converts all NLP tasks into text-to-text tasks. T5 uses more hyperparameters and training data than BERT, but allows for the direct use of the same model, objective, training process, and decoding process for every task. However, T5 is a mere combination of BERT and GPT in which the correlation between the encoder and the decoder enables contextualized output embedding as well as a self-attention mechanism that allows each word to attend differently to inputs based on different attention mechanisms.\\

More recently, transformer decoder-based generative language models, such as GPT-3 \cite{gpt3}, ChatGPT \cite{laskaracl}, GPT-4 \cite{gpt4}, PaLM \cite{palm1,palm2}, and LLaMA \cite{llama1,llama2}, have drawn considerable attention owing to their impressive capability to solve a wide range of tasks without requiring any task-specific fine-tuning. These decoder-based auto-regressive large transformer language models are pretrained on a large amount of self-supervised data, followed by the use of techniques of alignment, like reinforcement learning, based on human feedback to generate human-like responses. Although these LLMs have demonstrated impressive zero-shot performance, using them in real-world scenarios is challenging due to their consumption of a large amount of computational resources even during inference. Moreover, they can generate responses based only on their pretraining knowledge. This means that they cannot access any information beyond their pretraining timeline. To address
these limitations, retrieval-augmented generation has drawn a considerable amount of interest \cite{Qia}. In this approach, LLMs can retrieve information from external corpora to augment the context of the input. This helps them effectively answer even time-sensitive queries.  Specifically, external knowledge bases store vector embeddings of different documents. Thus, for a given query, the most relevant documents are retrieved based on the similarity between the query embedding and the document embeddings. Due to their effectiveness, embeddings generated by BERT-based models are widely used to retrieve the relevant information via vector similarity.\\

The relationship between RoBERTa \cite{RoBERTa}, ALBERT~\cite{Albert}, ELECTRA \cite{ELECTRA}, MacBERT \cite{MacBERT}, TK~\cite{Hofsttter2019, Hofsttter2020}, and T5 \cite{Raffel2019}, and the original BERT model ~~\cite{Devl} suggests that all these models are simply modified versions of the original BERT with adjustable objectives. We called these variants BERT-based models/approaches.
\section{BERT for Ad-hoc IR and Challenges}
\label{sec:application}
This section provides a comprehensive survey of state-of-the-art pretrained transformer encoders (i.e., BERT-based approaches) for the application of ad-hoc IR and the challenges to them. As BERT is the pioneer in pretrained transformer encoder models, we first briefly present the BERT model introduced by Devlin et al.~\cite{Devl} in Section~\ref{sec:introBERT}. For the purpose of analysis, all BERT-based IR approaches are divided into several categories, and the challenges to each are presented in Sections~\ref{sec:long}, \ref{aggr}, and \ref{sec:representation}. Section~\ref{sec:long} provides a comprehensive overview of BERT-based approaches that handle long documents. Considering the importance of integrating semantic information, Section~\ref{aggr} reviews recent BERT-based approaches from different perspectives on contextualization, namely, methods of aggregating long documents, ranking strategies, and weak supervision, in Section~\ref{aggregation}, Section~\ref{strategies}, and Section~\ref{weak}, respectively. We then provide a comparative analysis of these perspectives in Section~\ref{comparison}. Section~\ref{sec:representation} surveys BERT-based approaches that deal with effectiveness and efficiency by focusing on two aspects: reducing the duration of interactions in Section~\ref{dual}, and reducing the number of hyperparameters in Section~\ref{distillation}. In addition, Section~\ref{sec:representation} contains Subsection~\ref{training}, which summarizes BERT-based approaches and considers effective training strategies before a comparative study is provided in Section~\ref{experiment}. We also introduce a category of BERT-based term weight prediction in Section~\ref{sec:termweight}. Finally, we present a review of approaches to BERT-based query expansion and document expansion in Section~\ref{sec:qexpansion} and Section~\ref{sec:dexpansion}, respectively.
\begin{figure}[t!]
 \centering
 \includegraphics[width=0.9\textwidth]{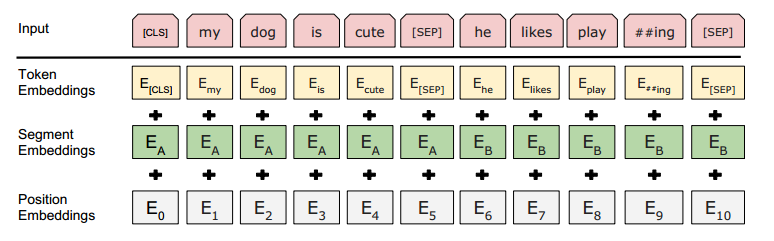}
\caption{The input representation of BERT consists of three parts: token embeddings, segment embeddings, and position embeddings. The special token [CLS] is always the first token in every sequence, and the special token [SEP] separates two sentences in it \cite{Devl}}.
 \label{f2}
\end{figure}

\subsection{Preliminary Overview of BERT}
\label{sec:introBERT}
BERT \cite{Devl} was introduced by Devlin et al. to develop a robust language model. Before the success of BERT, most of the state-of-the-art approaches (e.g., CoVe \cite{CoVe}, ELMo \cite{peter}, and GPT \cite{GPT}) were unidirectional or sequential, and thus failed to leverage contextual relationships across text inputs that led to a poor pretrained model. The bidirectional property of the BERT model allows it to learn the context of unlabeled text from both the left and the right sides such that it yields a good pretrained NLP model. BERT is essentially the encoder part of a transformer with the additional step of masking some of the input tokens and the additional objective of predicting the masked word based only on the context of the tokens surrounding it. The encoder receives each of the masked tokens that are conditioned on the remaining tokens in the sentence, where this renders the decoder redundant.\\

BERT provides both simple and complex models: $\rm BERT_{Base}$ (layers=12, hidden dimensions=768, self-attention heads=12, parameters=110M) and $\rm BERT_{Large}$  (layers=24, hidden dimensions=1024, self-attention heads=16, parameters=340M). Owing to its more complex architecture and a larger number of hyperparameters, $\rm BERT_{Large}$ outperforms $\rm BERT_{Base}$ under the same training and fine-tuning processes. The input representation of BERT consists of three parts: a WordPiece token embedding \cite{wordpiece} (i.e., the original vocabulary consists of individual symbols in the language, and the most frequent combinations of symbols are then iteratively added to the vocabulary), position embedding, and segment embedding (each token of the first sentence is denoted by $E_A$, and each token of the second sentence is denoted by $E_B$) (see Figure~\ref{f2} for details). There are two special tokens in BERT: the [CLS] token and the [SEP] token. The [CLS] token is always the first token of the input sequence for classification and the [SEP] token is the separator between sequences. The maximum length of an input containing the two special tokens is limited to 512.\\

BERT addresses the limitations and disadvantages of ELMo and OpenAI GPT by using the pretraining objectives of \textit{MLM} and \textit{NSP}. The \textit{MLM} randomly masks some words from the input texts during training and then predicts them from contextual information. During pretraining, $15\%$ of WordPiece  (i.e., a variant of byte pair encoding \cite{BPE} methods) \cite{wordpiece} tokens were chosen at random for prediction. Each chosen token had a $80\%$ chance of being replaced by the $[MASK]$ token, a $10\%$ chance of being replaced by a random token, and a $10\%$ chance of self-replacement. In addition, BERT uses \textit{NSP} to predict whether a pair of sentences is sequential (i.e., whether the second sentence is the next sentence in the original text). 

\subsection{Utilizing BERT for Handling Long Documents}
\label{sec:long}
As a pioneer pretrained language model, BERT~\cite{Devl} has drawn significant attention from the NLP community. Although BERT~\cite{Devl} is successful on short text-based applications (usually limited to 512 tokens), it fails to process long documents due to the limitation on the maximum length of the input sequence. To resolve this issue, state-of-the-art BERT-based methods rely on two categories of models: the aggregation-guided model and the block selection model.

State-of-the-art aggregation-guided methods ~\cite{Yan, Yil, Dep, LiM} focus on dividing a long document into small segments, e.g., sentences and overlapping passages, where the final document score is an aggregation of the top k scores retrieved by BERT and the original document score retrieved by a traditional model. Section~\ref{aggregation} provides a comprehensive review of aggregation-based approaches to BERT. It suggests that such approaches may suffer from a loss of important information that leads to problems related to time, memory, and energy consumption.\\
\begin{figure}[t!]
 \centering
 \includegraphics[width=0.7\textwidth]{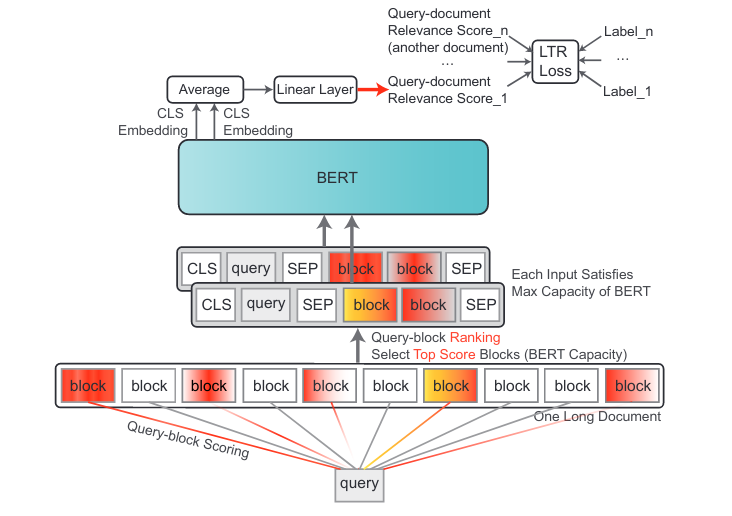}
\caption{Overview of the architecture of KeyBLD. \cite{LiM}. The KeyBLD model comprises four key components: block segmentation, block selection, query-blocks representation, and document ranking. The block segmentation phase divides documents into smaller blocks, while traditional IR models assign relevance scores to each block. The query-blocks representation step concatenates the query with the most relevant blocks, and this is the input to the BERT model for document re-ranking and improved retrieval performance.}
 \label{Key}
\vspace{-.5cm}
\end{figure}
Different from aggregation-guided methods, the block selection model (e.g., KeyBLD~\cite{LiM}) relies on selecting key blocks to form a short document that can then be processed by using a state-of-the-art BERT-based model. The idea of selecting key blocks with respect to the entire document by using the BERT model was proposed by Ding et al. in Cognize Long TeXts (CogLTX)~\cite{Ding}. Inspired by the cognitive theory proposed by Baddeley~\cite{badd}, CogLTX considers the analogy between BERT and human memory to implement long-term memory. The main step in CogLTX is MemRecall, the process of identifying the relevant text blocks by treating
the blocks as episodic memories. MemRecall imitates the working memory in terms of competitive retrieval, rehearsal, and decay to facilitate multi-step reasoning. Another BERT model, called judge, is introduced to CogLTX to score the relevance of the blocks and is trained jointly with the original BERT reasoner. Although the use of CogLTX for IR applications has not been explored, it works as a basic building block for KeyBLD~\cite{LiM}, which is used in IR applications. In contrast to CogLTX,  KeyBLD~\cite{LiM}  first selects the key blocks of a long document by local query--block preranking, and then aggregates a few blocks to form a short document that can be processed by BERT for ad-hoc IR. Figure~\ref{Key} illustrates the basic architecture of KeyBLD. The experimental analysis of KeyBLD~\cite{LiM} indicates that it outperforms the relevant state-of-the-art approaches and is faster than them.
\subsection{Utilizing BERT for Integrating Semantic Information}
\label{aggr}
The bidirectional property of BERT-based models enables the learning of the context (i.e., semantic information) of unlabeled texts from both the left and the right sides, and lays the foundation for obtaining contextualized embeddings. We first briefly discuss the use of BERT-based models for aggregating long documents in Section~\ref{aggregation} and then present ranking strategies in Section~\ref{strategies}. We then cover techniques of weak supervision that leverage BERT-like models in Section~\ref{weak}. Finally, we provide an extended comparison and analyses based on experiments in Section~\ref{comparison}.
\subsubsection{$\textbf{Aggregation Methods for Long Documents}$}
\label{aggregation}
A number of state-of-the-art approaches~\cite{Yan,Yil,Dep,LiM} can be considered to be aggregation techniques that combine shorter segments of texts to handle long documents. Aggregation-guided BERT models for IR can be divided into three groups, sentence-level score aggregation, passage-level score aggregation, and passage-level representation aggregation, as shown in Table~\ref{agg_t}. This section briefly discusses aggregation-guided BERT-based approaches as a subclass of BERT-based models for integrating semantic information, while the previous section (Section~\ref{sec:long}) considered aggregation-guided methods and block selection models for handling long documents.

\begin{figure}[t!]
 \centering
 \includegraphics[width=0.7\textwidth]{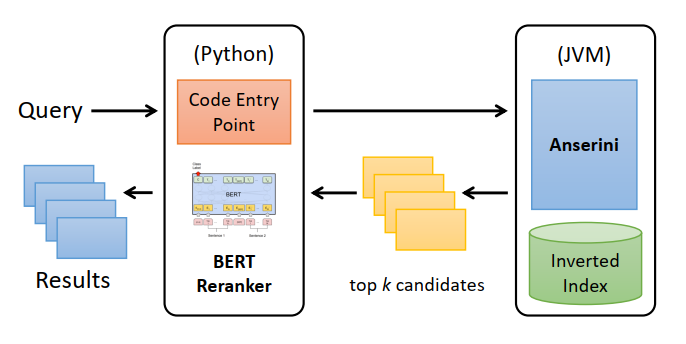}
\caption{The architecture of Birch showcases seamless integration between Python and the Java Virtual Machine (JVM), thus enabling the efficient utilization of neural networks like BERT. The main code-entry point implemented in Python leverages the capabilities of the JVM to facilitate document retrieval through the Lucene search library. The top k candidates retrieved from JVM’s Anserini are then input to BERT for a secondary ranking process \cite{Yil}.}
 \label{birch}
\end{figure}

$\bullet$ Sentence-level score aggregation: The state-of-the-art~\cite{Yan,Yil} approaches of this sort divide the document into a set of sentences, where the final document score is a linear aggregation of the top k scores retrieved by BERT and the original document score retrieved by the traditional model. Yang et al.~\cite{Yan} used the sentence-level model to apply BERT to document retrieval and divided the document into a set of sentences by combining the BERT score with the BM25 (Anserini \cite{anserinilin}) score to obtain the final score of document retrieval. Specifically, the final document score was a linear aggregation of the top k scores retrieved by BERT and the original document score retrieved by the traditional model. Yilmaz et al.~\cite{Yan} proposed a system called $Birch$ \cite{Yil}, which is similar to the proposal in \cite{Yan}, that is shown in Figure~\ref{birch}. Both these studies (\cite{Yil}, \cite{Yan}) reported experiments on two datasets: the TREC 2011-2014 Microblog and Robust04. The lengths of the documents in the TREC Microblog dataset fitted the BERT model well. Every document in the Robust04 dataset was split into a set of sentences. The results on the TREC Microblog dataset showed that BERT outperformed previously developed neural ranking models and demonstrated its suitability for passage ranking. The experiments on the Robust04 dataset involved a linear aggregation of the scores of the top three sentences and the original document score as the final document score. The results of Birch show that fine-tuning across multiple datasets can help obtain a better fit with the target dataset. For example, Birch ($\rm MS\rightarrow MB$) achieved the best results and outperformed Birch fine-tuned on MS MARCO, while Birch ($\rm MS\rightarrow MB$) was first fine-tuned on MS MARCO (MS) and then on the TREC Microblog (MB) dataset. However, the sentence-level score aggregation-based BERT suffers from the loss of important information as well as relevant judgment.
\begin{table}[t!]
\caption{Aggregation methods for handling long documents}
\label{Tab01}
\centering
\begin{tabular}{c|c|p{6cm}}
\toprule
Type & Model & Aggregation Methods\\
\midrule
Sentence-level score aggregation &\cite{Yan}, \cite{Yil}  & Top-3\\
\midrule
\multirow{3}{*}{Passage-level score aggregation} &\cite{Dep}&FirstP, MaxP, SumP \\
&\cite{Zhang2021}&AvgP \\
&\cite{Hofsttter2021}&TopKP\\
\midrule
\multirow{3}{*}{Passage-level representation aggregation} &\cite{Sea}&AvgP \\
&\cite{Li2020}&$\rm PARADE_{Avg}$, $\rm PARADE_{Sum}$, $\rm PARADE_{Max}$, $\rm PARADE_{Attn}$, $\rm PARADE_{CNN}$, $\rm PARADE_{Transformer}$ \\
\bottomrule
\end{tabular}
\label{agg_t}
\vspace{-.3cm}
\end{table}

$\bullet$ Passage-level score aggregation: The state-of-the-art approaches~\cite{Dep, Zhang2021, Hofsttter2021} of this kind rely on segmenting long documents into small overlapping passages. These passages are formed by considering the passage length and the stride length to prevent passages of a long document from becoming independent of one another. Hence, document-level retrieval can be realized by aggregating passage scores \cite{Dep, Zhang2021, Hofsttter2021}. Dai and Callan proposed different aggregation scores that split a long document into overlapping passages of length 150, i.e., a 150-word sliding window with a stride of 75 words \cite{Dep}. Any of three approaches can be used to obtain the final document-level retrieval score: 1) BERT-FirstP: by selecting the score of the first passage; 2) BERT-MaxP: by selecting the maximum passage score; and 3) BERT-SumP: by taking the sum of all passage scores. Zhang et al. proposed AvgP \cite{Zhang2021}, which takes the average score of all passages as the final document retrieval score. MaxP and TopKP must obtain all passage relevance scores via an expensive pretrained model, e.g., BERT, and select the maximum relevant passage or the top k passages. To simplify the selection process, Hofst{\"a}tter et al. selected the top k passages through a lightweight and fast model \cite{Hofsttter2021} called the ESM (efficient student model), and estimated the selected passages by using an expensive model. Despite its success, this strategy may still cause problems related to time, memory, and energy consumption. Furthermore, considering passages irrelevant to the given query may introduce noise to the final representation and limit the identification of long-distance dependencies between the relevant tokens.

$\bullet$ Passage-level representation aggregation: State-of-the-art approaches~\cite{Sea, Li2020} that fall into this category rely on aggregating the passage representations rather than the passage scores. Li et al. proposed passage representation aggregation for document reranking (PARADE) \cite{Li2020}. It uses six approaches to aggregate passage-level signals for a long document: 1) $\rm PARADE_{Avg}$: adopting average pooling for the passage representations of a long document to obtain the final document representation, where this is similar to CEDR \cite{Sea}; 2) $\rm PARADE_{Sum}$: performing additive pooling across passage representations; 3) $\rm PARADE_{Max}$: using element-wise max pooling for all passage representations to obtain the representation of the final document; 4) $\rm PARADE_{Attn}$: adopting a feedforward network to generate an attention weight for each passage; 5) $\rm PARADE_{CNN}$: generating pairs of passage representations using several convolutional neural networks (CNNs) until only one remains; and 6) $\rm PARADE_{Transformer}$: aggregating passage representations through two randomly initialized transformer encoders. To obtain context-aware passage-level signals, Wu et al. introduced the passage-level cumulative gain model (PCGM) to rank documents \cite{Pcg}. In their model, BERT was first applied to obtain passage-level [CLS] representations, following which they were aggregated via LSTM \cite{Graves2005}.

\subsubsection{$\textbf{Ranking Strategies via BERT}$}
\label{strategies}
\begin{figure}[t!]
 \centering
 \includegraphics[width=0.8\textwidth]{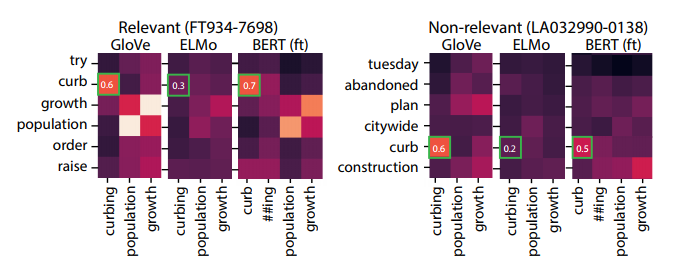}
\caption{Comparison of GloVe word embeddings, ELMo representations (layer 2), and fine-tuned BERT representations (layer 5) for relevant and irrelevant documents in relation to a given query \cite{Sea}. Lighter colors represent higher similarity scores.}
 \label{cedr}
\end{figure}

To investigate how pretrained transformer encoder-based contextualized language models can be utilized for ad-hoc document ranking, researchers have attempted to retrieve documents via various ranking strategies. As shown in \cite{Pas, Yil, Yan}, BERT-like models have a better capability of representation than state-of-the-art non-BERT-based IR approaches~\cite{Cro,Dss,Cnn,drmm,Srnn,Knr} for ad-hoc document ranking. MacAvaney et al. introduced contextualized embeddings for document ranking (CEDR)  \cite{Sea} that combines BERT embeddings with neural ranking models to enhance term representations and improve the performance of neural ranking models. They combined the contextualized word embeddings of BERT and ELMo with certain neural ranking models, including the position-aware deep model for relevance matching in IR: PACRR \cite{Pac}, KNRM \cite{Knr}, and DRMM \cite{drmm}. The BERT model can simultaneously encode multiple text segments by utilizing the two meta-tokens ([SEP] and [CLS]) and text segment embeddings (e.g., Segments A and B). [SEP] acts as a separator of each segment while
[CLS] is used to make contextual judgments about the text pairs. More precisely,  [CLS] is used during training to predict whether two segments are sequential, e.g., whether Segment A immediately precedes Segment B in the original text. Thus, incorporating the [CLS] token into the neural network allows the neural rankers to benefit from deep semantic information from BERT in addition to individual contextualized token matches. Incorporating the [CLS] token into the ranking models is a two-step process. First, the model produces relevance scores for each query term based on similarity matrices. Then, for models using a dense combination, CEDR concatenates the [CLS] vector with the signals of the model. For models that sum query term scores, such as the DRMM \cite{drmm}, CEDR shows that using the [CLS] vector in the dense calculation is effective. To better understand the contributions of contextualized word embeddings to the model, the experimental evaluation of CEDR involves visual comparison of GloVe word embeddings, ELMo representations (layer 2), and fine-tuned BERT representations (layer 5) as shown in Figure~\ref{cedr}. For the query ``curbing population growth'' in this figure, the term "curb" that appears in both relevant and non-relevant documents has two distinct senses. However, GloVe predicts the same similarity score for both candidates. By contrast, both ELMO and BERT assign a higher similarity score to the appropriate sense of the term in response to the query, demonstrating the capabilities of semantic understanding of contextualized models. Their performance on the Robust04 and WebTrack 2012-14 datasets demonstrated that the neural ranking models that used fine-tuned BERT representations (CEDR-PACRR, CEDR-KNRM, and CEDR-DRMM) outperformed the others. This verified the influence of contextualized embeddings on the performance of neural ranking models.

Qiao et al. \cite{Qia} systematically analyzed the [CLS] embedding in different layers of BERT. In their work, three interaction-focused models based on the $\rm BERT_{Large}$ model were introduced: $\rm BERT_{Last\mbox{-}Int}$, $\rm BERT_{Mult\mbox{-}Int}$, and $\rm BERT_{Term\mbox{-}Trans}$. $\rm BERT_{Last\mbox{-}Int}$ extracts features from the last layer of the [CLS] embedding and then combines them with a weight parameter. Inspired by $\rm BERT_{Last\mbox{-}Int}$, $\rm BERT_{Mult\mbox{-}Int}$ extracts features from the [CLS] embeddings of all layers. Each layer's [CLS] embedding is first combined with a weight parameter, and the features of all layers are then integrated. $\rm BERT_{Term\mbox{-}Trans}$ produces a transition matrix for the query and the document, and then integrates the transition matrices from all layers via mean pooling. Qiao et al. \cite{Qia} conducted experiments on two datasets: MS MARCO and ClueWeb09-B. The results showed that the three interaction-based BERT models delivered varying results on different datasets. $\rm BERT_{Last\mbox{-}Int}$ delivered better performance than the other two models on the MS MARCO dataset, and $\rm BERT_{Mult\mbox{-}Int}$ delivered the worst performance in terms of MRR@10. However, $\rm BERT_{Mult\mbox{-}Int}$ outperformed the other two methods on CluWeb09-B in terms of NDCG@20 and ERR@20.
\begin{figure}[t!]
 \centering
 \includegraphics[width=0.9\textwidth]{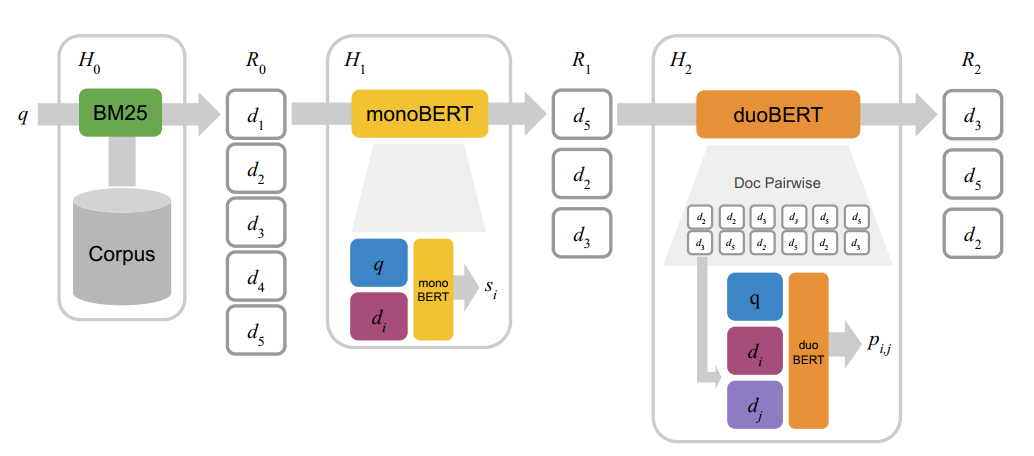}
\caption{Architecture of the multi-stage ranking model proposed by Nogueira et al. \cite{Nog}. The process consists of three stages, denoted by $H_0$, $H_1$, and $H_2$. The first stage, $H_0$, produces an initial ranking list retrieved by a simple traditional model for the entire corpus. The second stage, $H_1$, $\rm monoBERT$, serves as a reranker to obtain a more precise ranking. In the third stage, $\rm duoBERT$ adopts pairwise learning to further enhance the quality of the ranking list.}
\label{multi}
\end{figure}

All the important models discussed above examine correlations between documents and a query to obtain a final, sorted list related to the query. This raises the following question: If all elements of a previously sorted list are fed in pairs into BERT for comparison and a second sorted list is thus obtained, is the latter list consistent with the former list? Nogueira et al. provided an answer in \cite{Nog}. They used three stages, denoted by $H_0$, $H_1$, and $H_2$, in an iterative process (i.e., the output of each stage served as the input to the next stage). The first stage, $H_0$, produced an initial ranking retrieved by a simple traditional model for the entire corpus that was denoted by $R_0$ (i.e., the top $k_0$ candidate documents). In the second stage, $H_1$, $\rm monoBERT$ \cite{Nog}, served as a reranker to obtain a more precise ranking $R_1$ from $R_0$. Finally, the top $k_1$ candidate documents of $R_1$ were fed into $\rm duoBERT$ \cite{Nog} in the third stage $H_2$. Furthermore, $\rm duoBERT$ used a triple $(q,d_i,d_j)$ as input to produce the probability $p_{i,j}$ used to assess the degree to which candidate $d_i$ was more relevant than $d_j$, where $d_i,d_j\in R_1$. The architecture of this system is illustrated in Figure~\ref{multi}. The duoBERT model can be characterized as a pairwise approach and monoBERT as a pointwise approach in this framework. The authors conducted experiments on the MS MARCO and TREC CAR datasets, and represented the results in terms of the mean reciprocal rank (MRR) and the mean average precision (MAP) during different stages of retrieval. The experimental findings indicated that a multi-stage architecture offers more advantages in generating an accurate ranking list than a single-stage architecture. The primary limitation of the single-stage architecture lies in its inability to learn the relationships between documents. By contrast, the multi-stage architecture addresses this challenge by framing the problem of document ranking as a comparison of different documents. This approach is crucial for achieving a high-quality ranking list. While $monoBERT$ and $duoBERT$ apply different ranking strategies in different stages, duoBERT can more precisely find the most relevant document in a pair but incurs a higher computational cost.

\subsubsection{$\textbf{Weak Supervision via BERT}$}
\label{weak}
While supervised learning technology has achieved great success, it has come at the cost of expensive data annotation. Like many deep learning architectures, the data-hungry nature of BERT poses a significant challenge to accurate retrieval. Dai and Callan~\cite{Dep} found that when BERT was fine-tuned on user clicks from Bing, its accuracy on TREC Web Tracks improved by 16\% in comparison with the scenario when only masked LM-based pretraining was used. Weakly supervised pretrained models are trained on extensive amounts of data with limited or incomplete labeling information, leading them to acquire more general features and semantic representations. Consequently, these models exhibit impressive performance on new domains or tasks. On the contrary, the efficacy of a supervised model on a specific task may be constrained by the domain of its training data.

\begin{figure}[t!]
 \centering
 \includegraphics[width=0.8\textwidth]{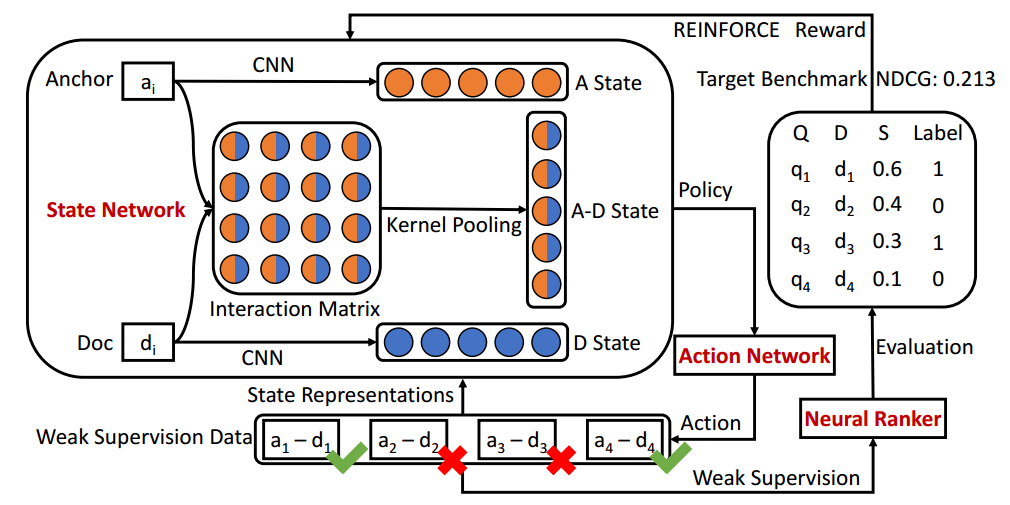}
\caption{The architecture of ReInfoSelect \cite{Sel} consists of a state network, an action network, and a reward. The state network represents the anchor-document (a-d) pair while the action network determines whether to retain the pair. The training reward is obtained from the ranker that has undergone training.}
 \label{wea}
\end{figure}

Although a number of weak supervision-based ad-hoc approaches to IR are available ~\cite{dehg,luo,zamani2018,zamani,laskarcl,laskarcoling,Rzhu20}, few studies~\cite{Sel,Wea} have considered a BERT architecture that relies on weak supervision. Zhang et al. proposed a reinforcement learning technique, $ReInfoSelect$ \cite{Sel}, that can guide ranking models to obtain more useful information. It consists of three components, a state network, an action network, and a training reward, as illustrated in Figure~\ref{wea}. The state network contains an a-d (anchor-document) pair that is fed into the action network to determine whether it should be used. $ReInfoSelect$ uses the $policy gradient$ approach to connect the state network to the action network to select useful a-d pairs for weak supervision. The remaining a-d pairs are fed into BERT to obtain a reward and then are returned to the state and action networks. Their normalized discounted cumulative gain values at a cut-off of 20 (as in NDCG@20) and ERR@20 on the ClueWeb09-B, Robust04, and ClueWeb12-B13 datasets, obtained by using both Conv-KNRM and BERT, were then compared. The results indicated good performance. Xu et al. \cite{Wea} proposed weak supervision for passage ranking (PR) via BERT. This model consists of three components: labeling functions, label aggregation, and supervised training. The labeling functions consist of four scoring functions: (1) BM25 score, (2) TF-IDF, (3) cosine similarity of the representation of a universal embedding \cite{universal}, and (4) cosine similarity of the last hidden layer. The most significant difference between the weakly supervised methods was in the manner in which the labels were obtained. To simplify the process of interaction, Hofst{\"a}tter et al. used the top k passages via a lightweight and fast selection model, the EST \cite{Hofsttter2021a}. They then used a computationally expensive model to estimate the relevance scores of the selected passages, rather than making relevance estimations for all passages of a document, to avoid wasting computational resources.

\subsubsection{$\textbf{Extended Analysis of Comparisons and Experiments}$}
\label{comparison}
In light of differences in the datasets and the evaluation indicators used in different studies,  their experimental comparison is provided in Table~\ref{Tab02}. We considered representative models and compared their values in terms of different evaluation metrics on the MS MARCO, TREC CAR, and Robust04 datasets. All the considered approaches~\cite{Pas,Sea,Nog,Qia,Sel} used consistent experimental settings that enabled us to draw a fair comparison. For example, for the MS MARCO dataset, all models in Table~\ref{Tab02} utilized the official BERT models that had been pretrained on the full Wikipedia website, while for the TREC CAR dataset, the BERT reranker was pretrained on half of the Wikipedia website to avoid the leakage of test data into the training data.
\begin{table}[h]
\caption{Comparison of empirical results of different methods in terms of MRR@10, MAP, and NDCG@20 on the MS MARCO, TREC CAR, and Robust04 datasets.}
\scriptsize
\renewcommand\arraystretch{1.2}
\label{Tab02}
\begin{tabular}{|p{3cm}|p{3.5cm}|p{1.8cm}|p{1.7cm}|p{1.6cm}|p{1.5cm}|}\hline
\hline
\multicolumn{2}{|c|}{Models} & \multicolumn{2}{c|}{MS MARCO dataset} & TREC CAR dataset & Robust04 dataset \\ \hline
Strategy & Approach &MRR@10 (Dev)  & MRR@10 (Eval) & MAP (Test) & NDCG@20 \\
\hline
\multirow{2}{*}{Traditional IR model} & BM25 (Microsoft Baseline) & 0.167 & 0.165 & 0.130 & 0.414 \\
                                      & BM25 (Anserini) & 0.187 & 0.190 & 0.153 & - \\
\hline
\multirow{2}{3cm}{\centering The different architectures of BERT} & $BERT_{Base}$\cite{Pas} & 0.347  & - & 0.310 & - \\
                                                     & $BERT_{Large}$ & 0.365  & 0.358 & 0.335 & - \\
\hline
\multicolumn{6}{|c|}{\leftline{\bf{Models based on $\mathbf{BERT_{Base}}$}}}\\
\hline
\multirow{6}{3cm}{\centering Passage-level representation aggregation via BERT}& PARADE-Avg \cite{Li2020} & - & - & - & 0.512\\
&PARADE-Sum  & - & - & - & 0.539\\
&PARADE-Max  & - & - & - & 0.544\\
&PARADE-Attn  & - & - & - & 0.527\\
&PARADE-CNN  & - & - & - & 0.563\\
&PARADE-Transformer  & - & - & - & 0.566\\
\hline
\multirow{3}{3cm}{\centering Combining BERT with neural ranking models} & CEDR-PACRR via BERT \cite{Sea} & - & - & - & 0.515\\
                                                           & CEDR-KNRM via BERT & - & - & - & 0.538 \\
                                                           & CEDR-DRMM via BERT & - & - & - & 0.526 \\
\hline
\multirow{3}{3cm}{\centering Weak supervision via BERT} &ReInfoSelect (Conv-KNRM)\cite{Sel} & - & - & - & 0.442\\
&ReInfoSelect (BERT) & - & - & - & 0.450 \\
\hline
\multicolumn{6}{|c|}{\leftline{\bf{Models based on $\mathbf{BERT_{Large}}$}}}\\
\hline
\multirow{6}{3cm}{\centering Multi-stage ranking via BERT} &(Anserini BM25) monoBERT \cite{Nog} & 0.372 & 0.365 & 0.348 & - \\
&monoBERT + $duoBERT_{MAX}$ & 0.326 & - & 0.326 & - \\
&monoBERT + $duoBERT_{MIN}$ & 0.379 & - & - & - \\
&monoBERT + $duoBERT_{SUM}$ & 0.382 & 0.370 & 0.369 & - \\
&monoBERT + $duoBERT_{BINARY}$ & 0.383 & - & 0.369 & - \\
&monoBERT + $duoBERT_{SUM}$ + TCP (target corpus pretraining) & 0.390 & 0.379 & - & - \\
\hline
\multirow{3}{3cm}{\centering Constructing various contextualized embeddings via BERT} &BERT (Last-Int) \cite{Qia} & 0.337 & 0.359 & - & - \\
&BERT (Mult-Int) & 0.306 & 0.329 & - & -\\
&BERT (Term-Trans) & 0.331 & 0.356 & - & -\\
\hline
\end{tabular}
\end{table}

In Table~\ref{Tab02}, both $\rm BERT_{Base}$ and $\rm BERT_{Large}$ serve as the reranker to rerank candidate documents retrieved by the traditional model, e.g., BM25. The above comparison yields three main conclusions:

$\bullet$ The MS MARCO dataset has an average of one relevant passage per query. To estimate the performance of the models on MS MARCO, the researchers used the mean reciprocal rank of the first correct answer. Because Anserini \cite{anserinilin} aims to connect academic research on IR with the development of search applications, the initial candidate list produced by Anserini's implementation of the BM25 algorithm exhibited a notable improvement of 12\% over the BM25 implementation by Microsoft on the MS MARCO dataset. This distinction had a noticeable impact on the final retrieval performance, specifically when comparing monoBERT with the BERT models introduced by Nogueira et al. \cite{Pas}. In this context, monoBERT utilizes BERT as the binary relevance classifier. This has also been illustrated by Lin et al. \cite{Lin2021}. Moreover, monoBERT + $duoBERT_{BINARY}$ and monoBERT + $duoBERT_{SUM}$ outperformed monoBERT + $duoBERT_{MIN}$ on the development set. By using the $TCP$, monoBERT + $duoBERT_{SUM}$ achieved the best performance. monoBERT + $duoBERT_{MAX}$ delivered the worst performance on the development set, $6\%$ worse than that of $BERT_{Base}$. In addition, $BERT_{Base}$ outperformed $BERT_{Last\mbox{-}Int}$, $BERT_{Multi\mbox{-}Int}$, and $BERT_{Term\mbox{-}Trans}$. Furthermore, Nogueira et al. applied the $\rm [CLS]$ embedding of the last layer, which is similar to that of $BERT_{\rm Last\mbox{-}Int}$. However, the variable gap in performance among $BERT_{Base}$, $BERT_{Last\mbox{-}Int}$, $BERT_{Multi\mbox{-}Int}$, and $BERT_{Term\mbox{-}Trans}$ might have stemmed from the variable batch size. In general, it seems that applying BERT to obtain the correlation between candidate documents yields a significant performance improvement, which in turn yields a novel area for sorting tasks in the field of IR.

$\bullet$ Similarly, monoBERT + $duoBERT_{BINARY}$ and monoBERT + $duoBERT_{SUM}$ delivered the best performance on the TREC CAR dataset, while monoBERT + $duoBERT_{MAX}$ performed the worst, about $3\%$ lower than $BERT_{Large}$.

$\bullet$ The original DRMM was used to compute the cosine distance between each word embedding of the query and all word embeddings of documents in the Robust04 dataset and then mapped them to a histogram. It can thus be considered a correlation model that omits location information. However, PACRR and KNRM can extract local contextualized information through their convolution kernels. Using the contextualized word embeddings of BERT, CEDR-KNRM achieved the best performance in terms of NDCG@20 on Robust04 and CEDR-PACRR performed slightly worse than CEDR-DRMM. The performance of PARADE-CNN and PARADE-Transformer was superior to that of the variants of CEDR, demonstrating that hierarchical aggregations can better preserve information that is beneficial to document-ranking. Unsurprisingly, ReInfoSelect (Conv-KNRM) and ReInfoSelect (BERT) performed slightly better than BM25.

\subsection{BERT for Balancing Effectiveness and Efficiency}
\label{sec:representation}
BERT models are more effective than previously proposed neural network ranking models in terms of document retrieval. However, $\rm BERT_{Base}$ has 110 million parameters and $\rm BERT_{Large}$ has 340 million parameters. Therefore, if the query and the text are fed simultaneously into BERT, its computational complexity is high, leading to a long computation time. To eliminate this deficiency, researchers have focused on two aspects: 1) reducing the interaction time between the query and the document, i.e., through dual-encoder-based models, and 2) reducing the number of hyperparameters of BERT, i.e., through distillation models. In light of these aspects, we discuss the state-of-the-art dual-encoder-based models, distillation models, and effective training strategies in Sections~\ref{dual}, \ref{distillation}, and ~\ref{training}, respectively. Finally, we provide an extended analysis of the relevant comparisons and experiments in Section~\ref{experiment}.

\subsubsection{$\textbf{Dual-encoder-based Models}$}
\label{dual}
Researchers have recently attempted to retrieve documents in a low-dimensional embedding space \cite{repbert, Xiong2020} through a dense retrieval model (DRM), the architecture of which is based on the representation-focused model. It maps queries and documents into low-dimensional embeddings via dual-encoders (i.e., the query encoder and the document encoder).

To achieve a better trade-off between effectiveness and efficiency, researchers have attempted to encode independent queries and documents to reduce the computational complexity of the stage involving their interaction. Qiao et al. proposed $\rm BERT(Rep)$ \cite{Qia}, which independently encodes the query and the document via BERT, and uses the [CLS] embeddings of the last layer as the representations of the query and the document. Then, relevance scores are calculated based on their cosine similarities. They used the MS MARCO and CluWeb09-B datasets for their experiments. ERR@10 was used for evaluation on the MS MARCO dataset while NDCG@20 and ERR@20 were used for evaluation on CluWeb09-B. BERT (Rep) was trained on ``Train Triples" provided by MS MARCO for its dataset.
\begin{figure}[t!]
 \centering
 \includegraphics[width=0.6\textwidth]{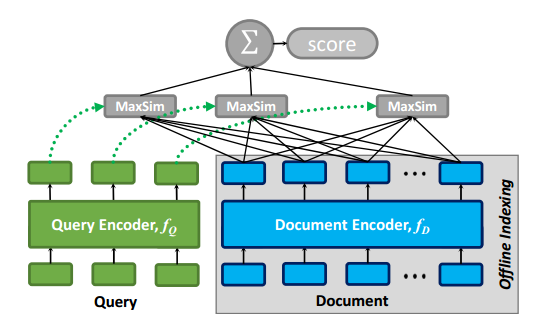}
\caption{ColBERT \cite{Col} independently encodes queries and documents via BERT, and can reduce the computational complexity of the model through late interaction.}
 \label{colb}
\end{figure}

BERT-based dense models of retrieval focus on dense passage retrieval \cite{sentence, repbert, Col, Luan2021, Wang2021}. Khattab \& Zaharia proposed a post-interaction architecture called ColBERT \cite{Col} that independently encodes queries and documents via BERT to reduce the computational complexity of the model ($n^{2}$ $\rightarrow$ $n$) and indexes the documents offline to reduce its time cost. $\rm ColBERT$ contains three components, a query encoder ($f_{Q}$), a document encoder ($f_{D}$), and an interaction machine, all of which are shown in Figure~\ref{colb}. ColBERT provides a novel approach to identifying the interactions between a query and a document. The query and the document are first encoded separately via BERT to store the word embeddings, and the similarity between each query embedding and all document embeddings is then calculated via an interaction machine to obtain the maximum score. Finally, all the maximum scores of the query--document token pairs are integrated to determine the relevance between the query and the document. The authors conducted experiments on the MS MARCO and the TREC CAR datasets, and provided a comparison with $\rm BERT_{Base}$ on the reranking and the end-to-end tasks in terms of MRR@10. $\rm BERT_{Base}$ was assessed in terms of MAP and MRR@10 values on the TREC CAR dataset. Based on the experimental results, $\rm ColBERT$ achieves better performance than BERT (Rep).

ColBERT \cite{Lin2020} was further improved by Lin et al. \cite{Lin2021} by using knowledge distillation. Reimers and Gurevych introduced Sentence-BERT \cite{sentence} for passage ranking to obtain sentence embeddings via the Siamese network. Similar to CEDR \cite{Sea} and $\rm PARADE_{Avg}$, RepBERT \cite{repbert} executes average pooling for all tokens in a passage to produce its final representation. DPR \cite{dpr} trains the query encoder and the document encoder on natural questions, Triviaqa (Trivia), web questions (WQ), CuratedTrec (TREC), and Squad V1.1.

\subsubsection{$\textbf{Distillation Models}$}
\label{distillation}

\begin{figure}[t!]
 \centering
 \includegraphics[width=1.0\textwidth]{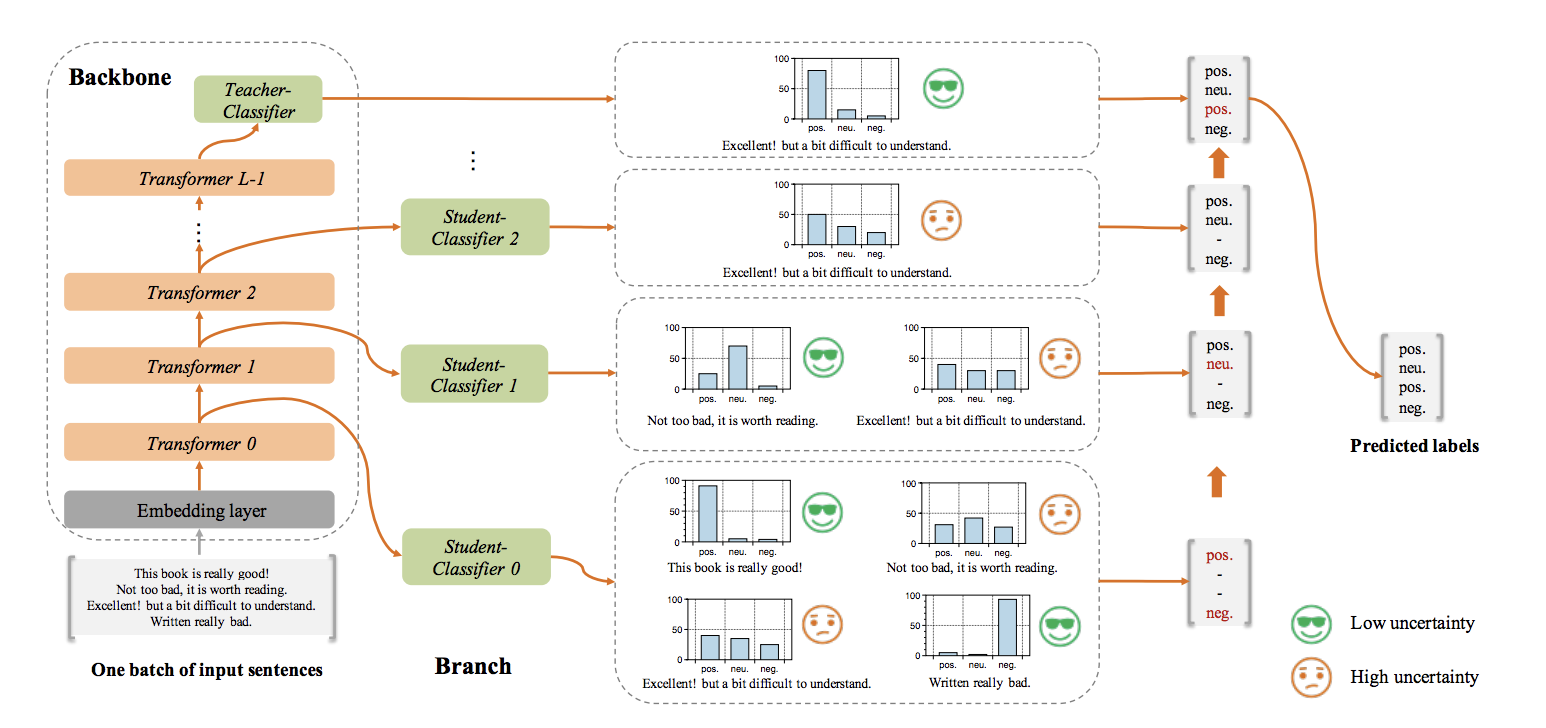}
\caption{The inference procedure of FastBERT \cite{Fastbert2020}. Given a batch of input sentences, the embedding layer translates it into an embedding sequence. Transformer0 and Student-classifier0 infer its labels as probability distributions while calculating uncertainty. }
 \label{fastbert}
\end{figure}

To address the limitations related to memory and the long time needed by state-of-the-art BERT models to make inferences, researchers have developed distillation models of BERT. Jiao et al. proposed TinyBERT \cite{tinybert} that uses transformer distillation, general distillation, and task-specific distillation. Transformer distillation relies on learning attention weights to enable the teacher (BERT) to transfer its linguistic knowledge to the student (TinyBERT), while the other two steps of distillation subsequently utilize the learned knowledge to perform task-specific distillation on augmented datasets. In contrast to TinyBERT, which uses distillation for building task-specific models, Sanh et al. proposed DistilBERT \cite{DistilBERT2019}, which uses knowledge distillation in pretraining to develop a lightweight, general-purpose, and pretrained model. By utilizing advanced optimization techniques and reducing the number of layers, DistilBERT \cite{DistilBERT2019} makes it possible to attain similar performance to that of larger models on many downstream tasks by using much smaller language models pretrained by using knowledge distillation. Unlike~\cite{tinybert,DistilBERT2019}, FastBERT \cite{Fastbert2020} uses a unique self-distillation mechanism during training and an adaptive mechanism for inferences to improve its efficiency without compromising its performance. Furthermore, FastBERT incorporates a dynamic layer adjustment mechanism to reduce computational complexity. This means that the model can adaptively modify the number of execution layers for each sample. For simpler samples, FastBERT utilizes fewer layers for prediction, while employing deeper layers for more intricate samples. Moreover, FastBERT incorporates a distinctive self-distillation processing mechanism that enables faster and more accurate results while minimizing changes to the model structure. The inferential process of FastBERT is illustrated in Figure~\ref{fastbert}. In contrast to the above distillation-based variants of BERT, Lan et al. proposed the ALBERT \cite{Albert} model, which has a significantly smaller size and improves the speed of inferences. Note that ALBERT mainly changes the model architecture through factorized embedding parameterization, cross-layer parameter sharing, and inter-sentence loss of coherence.

To reduce the temporal complexity of inference in the two-input case from quadratic to linear, TwinBERT \cite{Luw} uses twin-structured BERT-like encoders to encode the query and the document, respectively, and a crossing layer to combine the two embeddings and produce a similarity score. In contrast to BERT, the document representations in it are precomputed and stored via \textit{offline caching}. Due to the success of the knowledge distillation technique, the training of TwinBERT is based on the teacher-student framework.

\subsubsection{$\textbf{Effective Training Strategies}$}
\label{training}
Researchers have explored different training strategies for pretrained transformer encoder-based contextualized language models. For instance, Sentence-BERT (SBERT) \cite{sentence} uses Siamese and triple networks to generate contextual sentence embeddings. It uses cosine similarity or the Manhattan/Euclidean distance in the interaction stage to find semantically similar sentences. SBERT reduces the computational complexity of the system from $n^2$ to $n$ while ensuring accuracy. Approximate nearest-neighbor negative contrastive learning (ANCE) \cite{Xiong2020} constructs negative samples from the corpus via the approximate nearest neighbor (ANN) algorithm, where the total number of negative samples is dynamically changed based on the training samples used. It solves the issue of a difference in distributions between the training and the test datasets. The experimental results on MS MARCO and TREC DL showed that ANCE is superior to models that randomly sample negative values. The Topic-aware Sampling (TSA) \cite{Hofsttter2021} approach performs a clustering operation by using K-means for queries. Its training can be implemented according to the sampled topics. As a further improvement, Hofst\"{a}tter et al. used TSA-Balanced to sample the negative values in a batch via the maximum margin. The experimental results on MS MARCO and TREC DL showed that TSA-Balanced is superior to TSA and random sampling. Wang et al. \cite{Wangpr2021} followed the same pipeline that provides representative feedback embeddings via K-means clustering.

\subsubsection{$\textbf{Extended Analysis for Comparison and Experiments}$}
\label{experiment}
Research on representation-focused BERT models has mainly used the MS MARCO passage dataset. Due to its specificity (i.e., an average of one relevant passage per query), researchers use the mean reciprocal rank (MRR) to obtain the reciprocal rank of the first correct answer. We conducted an experimental comparison of representative models in terms of the MRR\@10 on the MS MARCO passage dataset (see Table~\ref{efff}).
\begin{table}[t!]
\caption{Comparison of empirical results in terms of MRR@10 on MS MARCO.}
\scriptsize
\renewcommand\arraystretch{1.2}
\label{efff}
\begin{tabular}{|p{4cm}|p{4cm}|p{2cm}|p{2cm}|}\hline
\multicolumn{2}{|c|}{Models} & \multicolumn{2}{c|}{MS MARCO Passage Retrieval Dataset} \\ \hline
Strategy & Approach &MRR@10 (Dev)  & MRR@10 (Eval)  \\
\hline
\multirow{2}{*}{Traditional IR model} & BM25 (Microsoft Baseline) & 0.167 & 0.165 \\
&BM25 (Anserini) & 0.187 & 0.190  \\
\hline
\multirow{3}{*}{Dual-encoder-based models} &RepBERT \cite{repbert}(over $\rm BERT_{Base}$) & 0.304 & 0.294 \\
&ColBERT (over $\rm BERT_{Base}$) \cite{Col} & 0.349 & 0.349 \\
&DPR (BM25 + Rand Neg)(over $\rm BERT_{Base}$) \cite{dpr} & 0.311 & -\\
\hline
\multirow{3}{*}{Effective training strategies} &ANCE (FirstP)\cite{Xiong2020}(over RoBERTa Base) & 0.330 & - \\
&TSA (Pairwise + In-Batch)\cite{Hofsttter2021}(over $\rm BERT_{Base}$) & 0.338 & -\\
&TSA-Balanced (Pairwise + In-Batch) & 0.340 & -\\
\hline

\multirow{3}{*}{Enhancing retrieval process with term weights} &DeepCT~\cite{Zhd} & 0.243 & -  \\
&EPIC + BM25 \cite{MacAvaney2020} & 0.273 & -\\
&SparTerm \cite{Baisparterm2020} & 0.275 & -\\
&SPLADE \cite{Formal2021} & 0.322 & -\\
&DeepImpact~\cite{Mallia2021} & 0.326 & -\\
&DeepImpact + ColBERT~\cite{Mallia2021} & 0.362 & - \\
\hline
\end{tabular}
\end{table}

 Table~\ref{efff} illustrates the superior performance of pretrained BERT models compared with sparse retrieval models, such as BM25. Among the representation-focused BERT models in dual-encoder-based architectures, ColBERT delivers the best performance on the MS MARCO dataset. It leverages token-level contextualized information to match queries and documents, thus enabling more precise targeting of the document content. On the contrary, RepBERT performs poorly on the MS MARCO dataset as it generates output-contextualized text embeddings through average pooling, which is not advantageous for capturing important representational information. Effective training strategies, specifically TSA-Balanced, outperform other models on the MS MARCO dataset, surpassing both TSA and ANCE. This finding highlights the significant role of negative sample selection in enhancing the capability of the model to discriminate between positive and negative samples. In conclusion, careful consideration of negative sample selection can significantly improve the training process and enhance the model's ability to distinguish between positive and negative instances.

\subsection{BERT-based Predictions of Term Weights }
\label{sec:termweight}
IR models consider three major factors in general: the $\textit{document length}$, $\textit{term frequency}$, and $\textit{term specificity}$~\cite{He2011, Huang2003,Zhao2011}. However, the importance of a term in a document also depends on the semantic information relating the query to the document~\cite{pan2023, Babashzadeh2013}.

\begin{figure}[t!]
 \centering
 \includegraphics[width=1.0\textwidth,  trim={0 0.0in 0 0},clip]{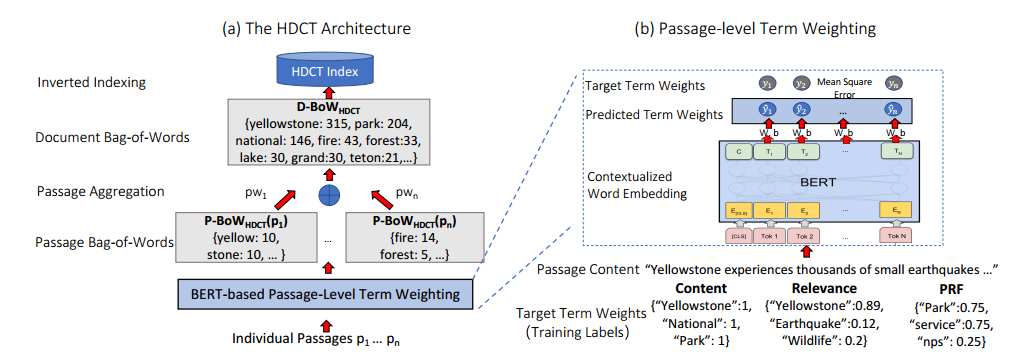}
\caption{Architecture of HDCT \cite{Zhd1}. The purpose of passage-level term weighting is to obtain the contextualized term weights for each passage. Moreover, HDCT integrates the contextualized term weights for all passages in the document for more accurate document ranking.}
 \label{hdct}
 
\end{figure}

To improve the efficiency of the retrieval stage, Dai and Callan proposed context-aware hierarchical document term weighting (HDCT) \cite{Zhd1} that provides a term-weighting framework for the retrieval stage in document ranking tasks, as illustrated in Figure~\ref{hdct}. Given a document $d$, HDCT first divides it into a set of passages. It produces contextualized word embeddings via BERT for tokens of every passage and then uses a linear layer to obtain a real number as weight. To ensure that integer values are obtained, HDCT uses two-digit precision to scale the predicted weight into an integer, such as the term frequency. The results of experiments showed that HDCT delivers significantly better performance than term frequency (tf).

Dai and Callan also proposed the DeepCT \cite {Zhd} framework that is similar to HDCT. Both use BERT to generate the contextual embedding for each token in the passage and obtain a term frequency-like integer instead of a term frequency value during the retrieval stage. The difference between the frameworks is that the former was designed to rank passages while the latter to rank documents. In contrast to these approaches, MacAvaney et al. proposed Expansion via Prediction of Importance with Contextualization (EPIC) \cite{MacAvaney2020}, which models the importance of query/document terms and executes document expansion. The reranking of passages by EPIC based on lexicon-grounded representations, generated from contextualized language models, helps improve its performance such that it outperforms DeepCT \cite {Zhd}, HDCT \cite{Zhd1}, and DocT5Query \cite{Nogueira2019} (as shown in Table~\ref{efff}). To address limitations in the training of the model as a per-token regression task in DeepCT and HDCT, Mallia et al. introduced DeepImpact \cite{Mallia2021} that can obtain independent term-based weights regardless of the co-occurrence of terms in the document. The major difference among the above methods is that DeepImpact leverages DocT5Query \cite{Nogueira2019} to enrich the original text via expansion terms.

Most of the above-mentioned studies sought to improve the capacity for dense representations of low-dimension queries (documents), whereas term-based sparse representations play an important role in industrial applications. To improve the sparse representations of Bag of Words (BoW) models (e.g., BM25), Xiong et al. introduced a framework called SparTerm \cite{Baisparterm2020} that consists of two components: an important predictor and a gating controller. The importance predictor can obtain a 30522D (i.e., the size of BERT's vocab.txt file) passage-wise distribution of importance for each passage by using a pretrained language model, and the gating controller provides expansion-enhanced vectors through binary word vectors of the passage. In a further refinement, Formal et al. proposed SPLADE \cite{Formal2021} on the basis of SparTerm. It adds a logarithmic function to the importance predictor to prevent the dominance of certain terms and smoothen the distribution of importance throughout the vocabulary. The major difference between SparTerm and SPLADE is in the loss function.

While these models require fine-tuning on task-specific datasets to achieve optimum performance, recently proposed LLMs like GPT-3 have demonstrated the capability of in-context learning via self-supervised pretraining on large amounts of data. Note that in-context learning works based on learning through analogies drawn from the given few-shot examples, thus making it possible for LLMs to solve tasks without requiring any task-specific fine-tuning. However, training only by using self-supervised objectives may still lead to poor performance on many tasks. To address this issue, the use of proximal policy optimization (PPO) with human feedback in a reinforcement learning (RL) framework has been recently proposed.  ChatGPT is the latest addition in this series that additionally uses dialog-based instructional data in the supervised and RL-based meta-training stages. ChatGPT has shown the ability to perform numerous tasks, and can outperform GPT-3. However, strong zero-shot performance across a wide range of tasks still does not help it achieve optimum performance on IR tasks.

The comparative experimental analysis reported in Table~\ref{efff} suggests that SparTerm and EPIC (based on BM25) deliver the worst performance and are inferior to SPLADE, which illustrates the importance of utilizing sparse representations. DeepImpact followed by a ColBERT re-ranker is more effective, where this underlines the importance of using independent term-based weighting.

\subsection{BERT-based Query Expansion}
\label{sec:qexpansion}

Document ranking in IR is primarily dependent on the appropriate understanding of the query terms to retrieve the relevant documents. Given the ability of BERT to capture contextual information based on the input text, it is expected that longer queries contain more information than shorter queries. Thus, the same collection of documents may result in a higher accuracy of document retrieval for longer queries. This situation has led to the emergence of the concept of query expansion~\cite{jimmy2013, Ye2011}.

\begin{figure}[t!]
 \centering
 \includegraphics[width=0.7\textwidth]{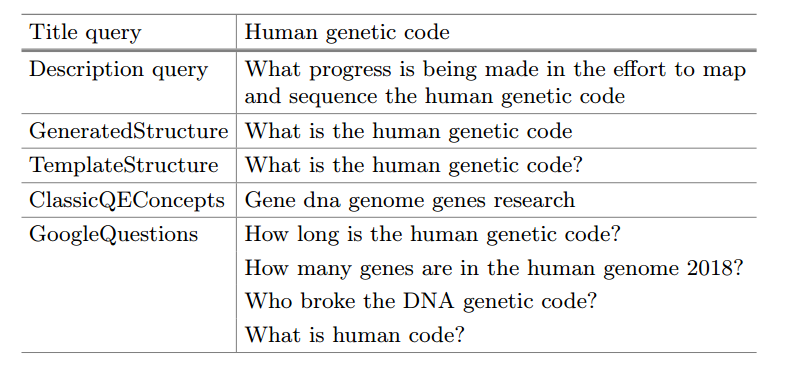}
\caption{A comparison of methods of query expansion \cite{Ret}.}
 \label{quep}
\end{figure}

To expand queries for reranking by BERT, Padaki et al. \cite{Ret} proposed three methods: $Expansion$ $with$ $structure$, $Expansion\ with\ concepts$, and $Hybrid$ expansion. The details of these approaches are as follows:
\begin{enumerate}
    \item $Expansion$ $with$ $structure$ produces a coherent sentence by adding structure words to it. The relevant study focused on two main approaches when adding a structure: $GeneratedStructure$ and $TemplatedStructure$. $GeneratedStructure$
copies keywords from the original query and adds question tokens to generate a new query without adding new concepts to the original query. $TemplatedStructure$ contains all the keywords of the original query in the new query and lets the structure expand until it reaches a maximum threshold.
\item $Expansion\
with\ concepts$ adds new concepts to the original query by using additional terms. $ClassicQEConcepts$ combines the relevance model RM3, which involves the classic pseudo-relevance feedback. A method of automatic local analysis is provided to improve retrieval performance without additional interaction to discover the relevant concepts.
\item The $Hybrid$ structure and concept expansions were implemented in a method called $GoogleQuestions$.
\end{enumerate}

Figure~\ref{quep} shows a comparison of these methods of query expansion. The experimental evaluation of these three methods~\cite {Ret} on Robust04 showed that ClassicQEConcepts based on discrete terms to expand the query delivered the worst performance, while FilteredGoogleQuestions based on structures and concepts delivered better performance on ``Query Title" and ``Query description."

\begin{figure}[t!]
\hfill
\subfigure[Process of chunk selection using the sliding window approach with BERT.]{\includegraphics[width=7.5cm]{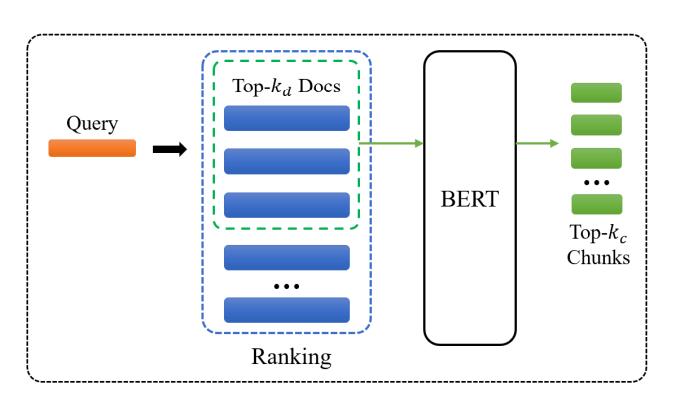}}
\hfill
\subfigure[Process of reranking documents by using selected chunks]{\includegraphics[width=7.5cm]{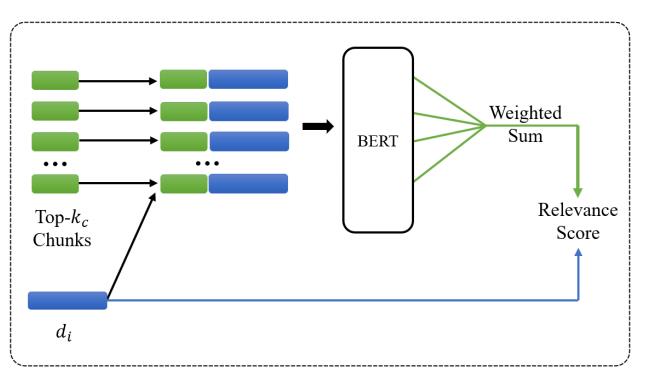}}
\hfill
\caption{Schematic illustration of BERT-QE proposed by Zheng et al. \cite{Qe}.}
\label{qe}
\end{figure}

Because BERT can obtain contextualized embeddings, researchers have explored the advantages of query expansion. Zheng et al. proposed the BERT-QE model in \cite{Qe}. It is a method of query expansion based on BERT for document reranking that uses important information from pseudo-relevance feedback (PRF). The BERT-QE model consists of three phases: (1) reranking the initial candidate documents retrieved by BM25 via BERT, (2) selecting fixed-length chunks for query expansion from the top-ranking documents obtained from phase one, and (3) reranking the documents with a combination of the original query and selected chunks. More details of the last two phases are shown in Figure~\ref{qe}. 
The results of experiments on the Robust04 and GOV2 datasets showed that BERT-QE outperformed $\rm BERT_{Large}$. Moreover, BERT-QE-LLL (BERT-Large during three phases) recorded gains of 2.5\%, 2.5\%, and 3.3\% on P@20, NDCG@20, and MAP@1k compared with $BERT_{Large}$, respectively. However, it had a 30\% higher computational cost than $\rm BERT_{Large}$. Moreover, all the above approaches failed to identify any new relevant document from the collection that had not been retrieved in the initial ranking.

Naseri et al. proposed contextualized embeddings for query expansion (CEQE) \cite{Naseri2021}, which is similar to BERT-QE.  It provides improvements based on a probabilistic language model combined with the PRF. It uses BERT to extract contextualized embeddings for the query as well as for the terms in the top-ranking feedback documents and then selects as expansion terms ones that are the closest to the query embeddings according to some similarity measure. In Robust04, the variations in the CEQE can yield improvements of 10.5\% and 19.9\% on P@20 and  MAP@1000 over the BM25, respectively. Unlike BERT-QE, CEQE can be used to generate a new query of terms for execution on a classical inverted index, but this comes at the cost of losing the contextual meaning of an expansion term (e.g., adding a polysemous word to the query can lead to topic drift).

To address this issue, recent state-of-the-art approaches~\cite{Wangpr2021,Yu2021} have exploited the advantages of using the PRF mechanism to improve query representations in scenarios of dense retrieval. The ColBERT-PRF~\cite{Wangpr2021} approach was the first to use PRF in a dense retrieval setting. As a robust mechanism of dense retrieval (a brief summary of ColBERT is provided in Section~\ref{dual}), ColBERT-PRF improves performance in terms of dense retrieval for query expansion in conjunction with PRF, which implements contextualized query expansion with dense retrieval, based on the pseudo-relevance set of documents identified from the first-pass retrieval. ColBERT-PRF utilizes K-Means clustering to extract representative feedback embeddings while ensuring that they can be used to discriminate among the representative embeddings. Subsequently, the representative embeddings are incorporated into the query representation. Extensive experimental evaluation of ColBERT-PRF on the TREC 2019 and 2020 supports their claims as it significantly outperforms the relevant approaches.

Following the same pipeline, Yu et al. proposed ANCE-PRF \cite{Yu2021}, which improves query representations directly from the pseudo-relevance labels by utilizing the query and the top retrieved documents from a dense retrieval model, ANCE~\cite{Xiong2020}. In this way, it learns to generate better query embeddings directly from the relevant labels such that the document index remains unchanged and this reduces overhead. The experimental analysis reported for ANCE-PRF~\cite{Yu2021} also indicated that under the same experimental settings, the margin of improvement obtained by using ANCE-PRF over its baseline ANCE was higher than that obtained by using strong models for dense retrievals, such as ColBERT or DistillBERT. 

\subsection{BERT-based Document Expansion}
\label{sec:dexpansion}
Inspired by methods of query expansion, researchers have explored expansions based on the documents themselves. Yan et al. proposed the $Doc2query$ method to expand the document before indexing \cite{Rod}. Given a document $d$, the $Doc2query$ model predicts a query related to it. The expanded document is a concatenation of the original document and the predicted query. In contrast to other methods of query expansion that expand query representations, $Doc2query$ aims to improve its capability of retrieval by expanding the representations of the documents. The retrieved candidate documents are fed into BERT for reranking. The results of experiments showed that BM25 yielded an improvement of more than 15\% in terms of the MAP and MRR@10 on both the TREC-CAR and the MS MARCO datasets by using Doc2query. In the same context, Yan et al.~\cite{Ids} proposed a document expansion-based retriever that first generates a query set for each passage, based on the pointer--generator model
\cite{Poi}, and then performs indexing. After the retrieval stage, the candidate documents are fed into BERT for reranking. Inspired by $StructBERT$ \cite{Str}, which incorporates the structural objectives of the word and the sentence to reconstruct their correct orders, the authors of this paper addressed the challenging task of predicting the sentences preceding and following a given sentence, instead of the original $NSP$ task. The results of comprehensive experimental evaluations in \cite{Ids} verified its superiority over the relevant state-of-the-art approaches in terms of both passage ranking and document ranking.
\begin{table}[htbp]
\centering
\caption{Statistical information on collections}
\label{compare}
\begin{tabular}{cccc}
\toprule
 Collection &  Query Creation & \# Queries & \# Documents \\
\midrule
 MS MARCO (Dev) & Manual & 6980 & 8.8M \\
 MS MARCO (Eval) & Manual & 6837 &8.8M \\
 \hline
  MS MARCO v2 (Dev1)& Manual & 3903  & 138M \\
 MS MARCO v2 (Dev2) & Manual & 4281  & 138M \\
 \hline
 TREC CAR & Automatic & 3.7M & 29M  \\
 \hline
 Robust04 & Manual& 250 & 0.5M  \\
 \hline
 ClueWeb09-B &  Automatic / Manual & 200 & 50M \\
 ClueWeb12-B13 &  Automatic / Manual & 100 & 52M\\
 \hline
 TREC 2011 &  Automatic / Manual & 50 & 16M \\
 TREC 2012 & Automatic / Manual& 60 & 16M \\
 TREC 2013 & Automatic / Manual & 60 & 240M \\
 TREC 2014 & Automatic / Manual & 55 & 240M \\
\bottomrule
\end{tabular}
\end{table}

\section{Resources for BERT-based Ranking Models}
\label{sec:Res}

In this section, we first present different datasets used in ad-hoc IR applications and then detail the resources available for BERT for IR, including MatchZoo \cite{matchzoo}, Anserini \cite{anserinilin}, Pyserini \cite{pyserini}, Birch \cite{Yil}, PyTerrier \cite{pyterrier}, Capreolus \cite{capreolus}, OpenNIR \cite{opennir}, and OpenMatch \cite{openmatch}, in Section~\ref{sec:resourse}.
\subsection{Datasets}
\label{sec:dataset}
Table~\ref{compare} provides information on the datasets that are typically used in ad-hoc IR applications. We describe them below.

\indent
$\bf{MS\ MARCO}$: The Microsoft Machine Reading COmprehension dataset \cite{Nguyen2016} consists of 8.8M passages for the ranking task. The training set contains 500K pairs of queries and the relevant documents, the development set contains 6,980 queries and an evaluation set with approximately 6,800 queries. In July 2021, Microsoft released an updated version of the MS MARCO collections called Version 2 (V2). Unlike the first version, in which the documents had been collected at different points in time, the passages and documents in MS MARCO V2 were created from the same underlying Bing web crawl. All the comparative statistics for MS MARCO V2 are shown in Table~\ref{compare}.

$\bf{TREC\ CAR}$: The TREC CAR dataset \cite{Dietz2017} consists of 29M documents extracted from English Wikipedia, and has five folds. The first four folds contain approximately 3M questions and the last fold contains 700K questions.

$\bf{Robust04}$: TREC 2004 Robust Track is a standard ad-hoc retrieval dataset \cite{Voorhees2004} containing 250 topics over a newswire corpus of approximately 500K documents.

$\bf{ClueWeb09\mbox{-}B}\footnote{\url{https://github.com/cdegroc/warc-clueweb}}$: ClueWeb09-B consists of 200 queries (i.e., Topics 1-50, 51-100, 101-150, 151-200) with relevance labels from
TREC Web Track 2009-2012, and contains 50M documents.

$\bf{ClueWeb12\mbox{-}B13}$: ClueWeb12-B13 contains 100 queries from TREC Web Track 2013-2014 and 52M webpages.

$\bf{TREC~Microblog}$: This consists of test tweet collections from the TREC Microblog Tracks 2011 to 2014 \cite{trec2014}. The salient targets of the track are social networks and real-time search. The TREC 2011 Microblog Tracks contains 50 queries and approximately 16 million tweets sampled from January 23 to February 8, 2011, from Twitter. For TREC 2012, the organizers used a new set of 60 topics and the same corpus as that of TREC 2011. TREC 2013 consists of 60 queries and approximately 240M tweets collected from Twitter from February 1 to March 31, 2013. TREC 2014 contains a set of 55 queries and the same collection as TREC 2013.

\subsection{ Resources for BERT-based Ranking Model }
\label{sec:resourse}
$\bf{MatchZoo}$\footnote{\url{https://github.com/NTMC-Community/MatchZoo-py}} is a text-matching toolkit that focuses on developing text-matching models. This includes the design and comparison of models. Examples of deep matching models include the DRMM, MatchPyramid \cite{Match}, MV-LSTM \cite{Mvls}, DUET \cite{Due}, ARC-I, ARC-II, and DSSM. MatchZoo uses a unified interface design for tasks including document retrieval, question-answering, session response ranking, and synonym recognition. It contains three main modules for data preprocessing, model building, and training and evaluation. Developed based on Keras, it supports TensorFlow, the Computational Network Toolkit (CNTK), and Theano, and works seamlessly on both CPUs and GPUs. MatchZoo focuses on the implementation of prevalent neural ranking models.

$\bf{Anserini}$\footnote{\url{http://anserini.io/}} provides wrappers and extensions based on core Lucene libraries to perform common research tasks by using intuitive APIs. The results of experiments on ad-hoc document retrieval have shown that BM25 (Anserini) is superior to BM25 because it performs better in first-stage retrieval.

$\bf{Pyserini}$\footnote{\url{https://github.com/castorini/pyserini}} is an easy-to-use Python toolkit for research on reproducible information retrieval with sparse and dense representations. The aim is to support the IR research community by making use of effective, reproducible, and easy-to-use first-stage retrieval in a multi-stage ranking architecture.$\bf{Birch}$\footnote{\url{https://github.com/castorini/birch}} is a framework for document ranking via BERT. It was written by using a combination of Python and Java Virtual Machine (JVM) code. The top k candidate documents retrieved by Anserini are fed into BERT for reranking.

$\bf{PyTerrier}$\footnote{\url{https://github.com/fennuDetudou/tudouNLP}} is a Python framework for conducting experiments on information retrieval based on Terrier\footnote{\url{http://terrier.org/}}, which provides an interface for downloading datasets and experimental functions. This allows for adequate comparisons between approaches to retrieval on the same queries and evaluation metrics.

$\bf{Capreolus}$\footnote{\url{https://github.com/capreolus-ir/capreolus}} is an NLP toolkit for ad-hoc document retrieval to enable mutual conversion between modules.
$\bf{OpenNIR}$\footnote{\url{https://github.com/Georgetown-IR-Lab/OpenNIR}} is a library for training and evaluating ranking tasks. It provides important information for fair comparisons, including a reranking threshold and the loss function. It is implemented on PyTerrier.

$\bf{OpenMatch}$\footnote{\url{https://github.com/thunlp/OpenMatch}} is an open-source package for IR that uses neural ranking models for matching and understanding deep text. It can be used for document retrieval, document reranking, and transfer learning across domains.

PyTerrier can provide interfaces for a detailed experimental setup to save the time needed to implement state-of-the-art indexing and retrieval functionalities. This enables the rapid development and evaluation of large-scale retrieval applications. Capreolus, OpenNIR, and OpenMatch are more suitable toolkits for drawing fair comparisons among state-of-the-art IR approaches by tuning the hyperparameters during training.

\section{Conclusions and Future Work}
\label{sec:conclusion}
In this paper, we provided a survey of the application of the BERT model to information retrieval. Because the BERT model is based on the pretrained transformer encoder architecture, we also studied other pretrained transformer encoders that are based on the BERT architecture. In our review, we summarized advanced methods based on the BERT architecture to deal with long documents, integrate semantic information, balance effectiveness and efficiency, predict term weights, and expand queries and documents. Furthermore, we analyzed the differences between approaches based on the BERT model to enable the reader to clearly understand the advantages and disadvantages of each. Due to inconsistencies in the datasets and evaluation metrics used for different models, we compared their published experimental results on benchmark datasets.

We presented resources for utilizing BERT for IR, including MatchZoo \cite{matchzoo}, Anserini \cite{anserinilin}, Pyserini \cite{pyserini}, Birch \cite{Yil}, PyTerrier \cite{pyterrier}, Capreolus \cite{capreolus}, OpenNIR \cite{opennir}, and OpenMatch \cite{openmatch}. We also discussed outstanding challenges to the use of BERT for IR according to two perspectives: 1) methods of aggregating long documents, and 2) approaches to balance effectiveness and efficiency. In addition, we emphasize the benefits of using encoder-based BERT models over recent LLMs such as ChatGPT, which are decoder-based and require significant computational power.

The experimental evaluations of state-of-the-art methods showed that their accuracy on different datasets in terms of MRR@10 was around 50\%. Hence, there is considerable room for improvement. A number of directions can be adopted in future research to this end. These include a novel combinatorial function to merge the contextualized embeddings of the BERT model with other neural ranking models for neural IR via multi-task optimization~\cite{Li22,amin22}, and reinforcement learning techniques~\cite{ghar22,gaur22} to transform weakly supervised learning into unsupervised learning to reduce the cost of training the data. Powerful generative large language models \cite{palm1,palm2,llama1,llama2,laskaracl,gpt4} can be leveraged across various stages of retrieval and ranking processes \cite{llmrankers, llmsearch}  and be applied for more applications (such as biomedicine and chemical IR) in the future~\cite{Huang2009,Yin2010, Lupu2009, Lupu20092}. Moreover, these models can be beneficial in enhancing techniques of query and document expansion \cite{queryexpansion}. Exploring the integration of Conversational Search Systems (CSS) \cite{kkey23} with multimodal IR presents a promising future direction for more dynamic, user-centric interactions and precise information retrieval through chat and voice technologies.

\section{Acknowledgments}
We gratefully appreciate all the anonymous reviewers and the associate editor for their valuable comments and constructive suggestions that greatly helped improve the quality of the paper. The authors would like to thank Prof. Xinhu Tu for his fruitful discussion when he visited our lab at York University. The authors also would like to thank Dr. Junmei Wang for her contribution at the early stage of research when she was a PhD candidate at Central China Normal University. All the work was done when the first author was a Ph.D. candidate. This research is supported by the research grant (RGPIN-2020-07157) from the Natural Science and Engineering Research Council (NSERC) of Canada, the York Research Chairs (YRC) program, and an Ontario Research Fund-Research Excellence (ORF-RE) award from the BRAIN Alliance.

\end{document}